\documentclass[12pt,english,letterpaper]{article}
\usepackage{amsmath,amssymb,dsfont,url,mathrsfs,nicefrac,xspace}
\usepackage[table,dvipsnames]{xcolor}
\usepackage[capposition=bottom]{floatrow}
\usepackage{lmodern}
\usepackage{mwe}
\usepackage[skins]{tcolorbox}
\usepackage{lipsum}
\usepackage{comment}
\definecolor{myred}{RGB}{213,94,0}
\definecolor{mygreen}{RGB}{0,158,115}
\definecolor{myblue}{rgb}{0,0,0.75}
\definecolor{bcblue}{RGB}{0,30,52}
\definecolor{posblue}{RGB}{0,80,158}
\definecolor{negred}{RGB}{185,30,50}
\definecolor{pivgreen}{RGB}{0,120,60}
\definecolor{defblue}{RGB}{0,70,170}
\definecolor{resultred}{RGB}{180,0,0}
\definecolor{alexblue}{RGB}{0,70,170}
\definecolor{shotaorange}{RGB}{230,120,20}
\usepackage[hyperfootnotes=false]{hyperref}
\hypersetup{
  colorlinks   = true,
  urlcolor     = myblue,
  linkcolor    = myblue,
  citecolor    = myblue,
  breaklinks=true,
  colorlinks=true,
}
\usepackage{tikz}
\usetikzlibrary{patterns, calc,shapes,decorations.pathreplacing,arrows.meta}
\usepackage{pgfplots}
\usepackage{subfig}
\usepackage{amsthm}
\makeatletter
\def\th@plain{%
\thm@notefont{}%
  \itshape
}
\def\th@definition{%
  \thm@notefont{}%
  \normalfont
}
\makeatother
\usepackage{mathtools}
\usepackage[english]{babel}
\usepackage{bm}
\usepackage{accents}
\usepackage{graphicx}
\usepackage[semicolon]{natbib}
\usepackage{booktabs}
\usepackage{enumitem}
\usepackage[framemethod=TikZ]{mdframed}

\theoremstyle{plain}
\newtheorem{theorem}{Theorem}
\newtheorem{lemma}{Lemma}
\newtheorem{proposition}{Proposition}
\newtheorem{corollary}{Corollary}
\newtheorem{claim}{Claim}
\theoremstyle{definition}
\newtheorem{definition}{Definition}
\newtheorem*{definition*}{Definition}
\newtheorem*{assumption*}{Assumption}
\newtheorem{assumption}{Assumption}
\newtheorem{conjecture*}{Conjecture}
\newtheorem{example}{Example}

\usepackage{url}
\usepackage{soul}

\usepackage[letterpaper]{geometry}
\usepackage{setspace}
\usepackage[bottom]{footmisc}
\newcommand{\type}{\ensuremath{c}}

\newcommand{\virtual}{\ensuremath{\gamma}}
\newcommand{\wtp}{\ensuremath{v}}

\newcommand{\typemeas}{\ensuremath{F}}
\newcommand{\typedens}{\ensuremath{f}}
\newcommand{\wtpmeas}{\ensuremath{G}}
\newcommand{\wtpdens}{\ensuremath{g}}
\newcommand{\rec}{recommendation algorithm\xspace}

\newcommand{\algo}{algorithm\xspace}

\newcommand{\alg}{\ensuremath{r}}

\newcommand{\E}{\ensuremath{\mathbb{E}}}
\newcommand{\expect}{\ensuremath{\mathbb{E}}}

\newcommand{\wtptr}{\ensuremath{\hat{v}}}
\newcommand{\popt}{\ensuremath{p^*}}

\newcommand{\dint}{\ensuremath{\mathrm{d}}}

\newcommand{\closing}{\ensuremath{\lozenge}}

\newcommand{\signal}{\ensuremath{s}}
\newcommand{\info}{\ensuremath{\mathcal{I}}}

\newcommand{\map}{\ensuremath{\mathcal{\pi}}}
\newcommand{\signals}{\ensuremath{S}}

\newcommand{\profit}{\ensuremath{\pi}}

\newcommand{\V}{\ensuremath{\tilde{\wtp}}}
\newcommand{\vst}{\ensuremath{y}}
\newcommand{\strategic}{pseudo\xspace}

\newcommand{\weight}{\ensuremath{\alpha}}

\newcommand{\argmax}{\mathop{\mathrm{argmax}}}

\newcommand{\vecsignal}{\ensuremath{\mathbf{\signal}}}
\newcommand{\vecwtp}{\ensuremath{\mathbf{\wtp}}}
\newcommand{\vectype}{\ensuremath{\mathbf{\type}}}
\newcommand{\vecp}{\ensuremath{\mathbf{p}}}

\newcommand{\BS}{\ensuremath{\mathrm{BS}}}
\newcommand{\SSurplus}{\ensuremath{\mathrm{SS}}}
\newcommand{\TS}{\ensuremath{\mathrm{TS}}}
\newcommand{\Wobj}{\ensuremath{\mathcal{W}}}
\newcommand{\win}{\ensuremath{\mathcal{A}}}
\newcommand{\Jset}{\ensuremath{\mathcal{J}}}
\newcommand{\Jzero}{\ensuremath{\mathcal{J}_0}}
\newcommand{\one}{\ensuremath{\mathds{1}}}

\newcommand{\sellerweight}{\ensuremath{\lambda}}

\newcommand{\Trade}{\mathcal T}

\newcommand{\ind}{\mathds{1}}
\providecommand{\payoffset}{\ensuremath{\mathcal{P}}}
\providecommand{\Trade}{\mathcal T}

\pgfplotsset{compat=1.17}
\usepgfplotslibrary{fillbetween}

\begin{document}
\hypersetup{pageanchor=false}
\title{Algorithmic Recommendations and Strategic Pricing\thanks{Ichihashi: Queen's University, Department of Economics, \href{mailto:shotaichihashi@gmail.com}{\texttt{shotaichihashi@gmail.com}}. Smolin: Toulouse School of Economics, University of Toulouse Capitole and CEPR, \href{mailto:alexey.v.smolin@gmail.com}{\texttt{alexey.v.smolin@gmail.com}}. We would like to thank Nemanja Antic, Heski Bar-Isaac, Dirk Bergemann, Alessandro Bonatti, Daniele Condorelli, Piero Gottardi, Emir Kamenica, Caio Lorecchio, Matthew Mitchell, Alessandro Pavan, Jacopo Perego, Nicola Persico, and Udayan Vaidya, and audiences at ESSET Gerzensee 2024, Berkeley-Columbia-Duke-Northwestern-MIT IO Theory Conference 2024, and other venues, for their valuable suggestions and discussions. The authors acknowledge support from the Economics of Digital Services initiative at the University of Pennsylvania and from the NET Institute. Ichihashi acknowledges funding from the Social Sciences and Humanities Research Council of Canada (SSHRC) through Insight Development Grant 430-2024-00031. Smolin acknowledges funding from the French National Research Agency (ANR) under the Investments for the Future (Investissements d'Avenir) program (grant ANR-17-EURE-0010) and through the Artificial and Natural Intelligence Toulouse Institute (ANITI). Javier Gonzalez-Morin provided excellent research assistance. This paper supersedes an earlier version circulated under the title ``Buyer-Optimal Algorithmic Recommendations.''}}
\author{Shota Ichihashi \and Alex Smolin}
\date{August 26, 2026}
\maketitle
\thispagestyle{empty}
\begin{abstract}
We study algorithmic recommendations in a market with a buyer, privately informed sellers, and an algorithm that can recommend a product based on product values and posted prices but cannot use transfers or force purchases. 
For objectives ranging from total surplus to buyer surplus,  algorithmic recommendations implement the same welfare outcomes as direct mechanisms. 
We study the effects of algorithmic recommendations on pricing, competition, the composition of trade, and market segmentation. 
Our results inform the operation of AI assistants and other technologies that mediate product discovery, attention, and pricing.
\end{abstract}

\newpage
\hypersetup{pageanchor=true}
\setcounter{page}{1}
\section{Introduction}
A fundamental feature of artificial intelligence and, more generally, of algorithms is that they are programmed in advance. As such, they endow their users with a degree of commitment power that those users may otherwise lack. In this paper, we study how this commitment power can affect market outcomes when it is bestowed upon buyers via algorithmic recommendations and sellers set prices strategically.\footnote{Such recommendations could come, for example, from a ChatGPT search or an AI shopping assistant on Amazon; see \cite{openai2025instantcheckout,openai2026productdiscovery,amazon2026alexashopping}.} 

We study a canonical market with a single buyer and multiple sellers. Each seller offers one product, privately observes her production cost, and sets a price strategically. 
The buyer is initially uninformed about the products' values and unaware of their existence. 
An algorithm, which can discover these values, recommends a product based on values and posted prices. 
Upon receiving a recommendation, the buyer updates her belief and decides whether to buy the recommended product at the offered price. If no recommendation is received, no trade occurs. 
Thus, the algorithm serves as a gatekeeper.

To evaluate such rules, we characterize the welfare frontier attainable by recommendation policies. In the main analysis, the designer puts weakly more weight on buyer surplus than on seller surplus. 
This class is natural for AI assistants and other recommendation tools whose long-run value depends more on buyer trust and subscription revenue than on seller commissions. 
It also covers platforms that value buyer participation and efficiency.

Our first main result is an equivalence between optimal recommendation algorithms and optimal direct mechanisms: optimal algorithms attain the same allocation, buyer surplus, and total surplus as optimal direct mechanisms. 
The equivalence is not mechanical. A direct mechanism can ask sellers to report costs, control allocation, and specify monetary transfers. 
In contrast, an algorithm only chooses whether and what to recommend based on the value and price of each product.
Nevertheless, the algorithm can use the equilibrium prices as messages to learn sellers' costs, and then use the revealed information to reproduce the allocation rule of the corresponding direct mechanism.

Specifically, the equivalence has two parts. 
First, any direct mechanism that satisfies incentive compatibility and the buyer-obedience constraints and gives each seller's highest-cost type zero interim profit can be implemented by some algorithm. 
Conversely, every equilibrium outcome of any recommendation algorithm satisfies the same restrictions. Second, for the objectives we study, the optimal direct mechanism automatically satisfies buyer obedience. Hence the requirement that the buyer be willing to follow recommendations imposes no loss at the optimum. 

The optimal allocation has a simple structure. The algorithm recommends a product that maximizes value net of weighted virtual cost whenever this maximum is positive. When the designer maximizes total surplus, the virtual cost equals the true cost and the allocation is efficient. When the designer puts more weight on buyer surplus, the algorithm treats costs as higher than their true levels to reflect information rents. 

The single-seller case illustrates the economic mechanism most transparently. The optimal algorithm recommends the product if and only if the posted price is below a function of the true value. 
This function differs from the identity function in a way that makes demand more price-sensitive, and is chosen so that, in equilibrium, trade occurs exactly when value exceeds the seller's virtual cost. 
With multiple sellers, the same force interacts with product selection: a seller's price affects not only whether her product is attractive to the buyer, but also whether it is attractive relative to other products, so high prices can additionally be punished by substitution toward rivals.

In \autoref{sec:implications}, we study the broader consequences of algorithmic recommendations.
First, we compare algorithmic recommendations with ordinary seller competition.
With continuously and independently distributed values and positive weight on buyer surplus, Bertrand competition without algorithmic recommendations is always improvable by some algorithm.
The algorithm improves on Bertrand competition by committing the buyer not to trade when trade would generate surplus but require high prices or large seller rents. 
At the same time, competition is still valuable and may partly substitute for the optimal algorithm.
In a common-value benchmark, the relative value of algorithmic recommendations decreases as the market becomes thicker.
Also, a merger weakly lowers the maximal value of every objective that puts positive weight on buyer surplus.

We then study how the algorithm's price-disciplining role changes which trades take place. 
With a single seller, compared to standard monopoly pricing,
the optimal algorithm substitutes low-cost, lower-value trades for high-cost, higher-value trades.
Thus, the algorithm replaces expensive surplus with cheaper surplus rather than merely transferring surplus for a fixed set of trades.
We extend this result to a parametric setup with multiple sellers.

Algorithmic recommendations also change the welfare analysis of third-degree price discrimination.
Suppose that each seller can base its price on some signal about the buyer's values.
The optimal algorithm tailored to such a situation neutralizes the average effects of market segmentation.
Specifically, price discrimination does not affect the product allocation, seller profits, or ex ante buyer surplus.
At the same time, finer segmentation makes prices more value-contingent: 
sellers charge higher prices in favorable segments and lower prices in unfavorable segments.
As a result, transaction prices become more dispersed without changing their mean.

Finally, we examine two extensions. 
First, if the buyer is already aware of the products, the algorithm has less control over the buyer's purchasing decisions. 
We show when the same price-disciplining logic survives and when the optimal algorithm must disclose additional information to deter off-path prices.
Second, we consider objectives that place more weight on seller surplus than on buyer surplus. 
Seller-oriented algorithms increase trade with higher-cost sellers to raise information rents.
Such algorithms may not select the lowest-cost seller and instead allow a higher-cost seller to transact at a high price.
Consequently, the buyer-obedience constraint, which imposes no loss for the buyer-oriented objectives, can now bind.
In this case, the equivalence with direct mechanisms can fail. 
Nevertheless, our main economic insight continues to hold: algorithmic recommendations expand and shape the set of market outcomes through their commitment power.

\medskip

\noindent \emph{Related literature.} Our paper is closest to work on buyer-optimal information design in markets with seller market power. \cite{roesler2017buyer} analyze buyer-optimal learning in a bilateral trade setting. \cite{dero2021} extend the analysis to multiproduct monopoly, \cite{bergemann2023bidder} to auctions, and \cite{condorelli2020information} characterize buyer-optimal distributions of values. We share the central insight that full information about value need not be optimal for the buyer. The key difference is that information in our model is price-contingent and the seller has private cost information. 
As a result, the recommendation rule is not only an information structure for the buyer but also a screening device for the seller. This feature leads to the implementation equivalence between recommendation algorithms and direct mechanisms.\footnote{The dependence of information on price may also arise from a worst-case analysis, as in \cite{libgober2021informational}, in which the buyer's information is chosen to minimize the seller's profits.}

In our study, we integrate Bayesian persuasion (cf. \cite{kage11}) with mechanism design (cf. \cite{baron1982regulating}). Several papers combine these frameworks in trade settings. The most closely related are \cite{yang2022selling}, \cite{bergemann2024data}, \cite{xuyang2024equivalence}, and \cite{bergemann2024digital}.\footnote{See also \cite{lee2021recommender}, \cite{bergemann2022screening}, and \cite{smolin2020disclosure}.} These papers also study informational intermediaries that process data and shape market outcomes. Our focus differs in two ways. First, these papers allow the intermediary to charge transfers and study revenue maximization, whereas we study general welfare objectives and an intermediary that cannot charge transfers. Second, in our model the central friction is seller private information. By contrast, \cite{bergemann2024data} and \cite{bergemann2024digital} focus on buyers' off-site purchase opportunities. Closest to us, \cite{yang2022selling} and \cite{xuyang2024equivalence} study revenue-maximizing intermediation when the intermediary has information about buyer values and faces a seller with private cost: the former characterizes optimal data-selling mechanisms, while the latter establishes outcome equivalence across retail and marketplace models and across recommendation and price-discrimination implementations. Our equivalence result is different: it shows that, for buyer-aligned objectives, a recommendation algorithm without transfers can be analyzed through the corresponding direct mechanism. 

Our market-segmentation results contribute to the literature on price discrimination and information design. In canonical segmentation models, information about buyer values changes the distribution of prices, allocations, and surplus; see, for example, \cite{bergemann2015limits} and \cite{haghpanah2023pareto}. In our setting, once the recommendation algorithm adjusts to the information available to sellers, price discrimination leaves the allocation, seller profits, and buyer surplus unchanged. What changes is the distribution of transaction prices across realized values: finer segmentation makes this distribution more dispersed. Thus the welfare effect of segmentation depends on whether buyer-side recommendation technology adjusts together with seller information.

Finally, our work  contributes to recent research strands in the economics of algorithmic decision-making. A  growing body of work, too large to survey here, examines algorithmic pricing, largely focusing on how it bolsters seller power.
Closer to this paper, \cite{calvano2025artificial} use simulations to study the impact of buyer recommender systems on market outcomes. They focus on a fixed class of algorithms that do not condition recommendations on prices. In contrast, we study general algorithms that can condition recommendations on prices.
Our results highlight that algorithmic recommendations can mitigate information asymmetries and counteract traditional price discrimination. As such, they can be viewed as an effective alternative to the joint use of an intermediary (see \cite{decarolis2021mad} for online advertising) or to a merger (see \cite{loertscher2022incomplete} for multifirm bargaining).\footnote{Based on our equivalence result, one can view the optimal recommendation algorithm as effectively enabling a buyer from the classic setting of \cite{myerson1983efficient} to commit to values and prices at which she would be purchasing a product. In this sense, we proceed in the opposite direction from the literature on limited commitment, which investigates how the inability to commit, typically on the part of a seller or a mechanism designer, affects equilibrium trade outcomes (e.g., \cite{mylovanov2014mechanism}, \cite{liu2019auctions}).}
\section{Model}\label{sec:model}

There is a single buyer with unit demand and $J \in \mathbb{N}$ sellers indexed by $j=1,\ldots,J$, each offering one product. 
The buyer's values for the products are
\[
\vecwtp\triangleq(\wtp_1,\ldots,\wtp_J)\in[0,1]^J,
\]
and are drawn from a joint distribution $\wtpmeas$ whose marginal on each $\wtp_j$ is not degenerate at zero.
Values may be arbitrarily correlated across products. 
Seller $j$'s cost $\type_j\in[0,1]$ is drawn from $\typemeas_j$, independently of other sellers' costs and of the value profile.
Assume each $\typemeas_j$ is continuous and has strictly positive density $\typedens_j$ on $(0,1)$. A dummy seller indexed by $j=0$, with $\wtp_0=0$ and $\type_0=0$, represents the buyer's outside option of not purchasing. Let
\[
\Jset\triangleq\{1,\ldots,J\}
\quad\text{and}\quad
\Jzero\triangleq\{0,1,\ldots,J\}.
\]

The buyer does not know her values and relies on external recommendations to make her purchases. A \emph{\rec} or simply \emph{\algo} is a function
\begin{equation*}
\alg:[0,1]^J\times\mathbb{R}_+^J\to \Delta(\Jzero),
\end{equation*}
where $\alg_j(\vecwtp,\vecp)$ is the probability that the algorithm recommends seller $j\in\Jzero$ at value profile $\vecwtp$ and price profile $\vecp$. Recommendation of seller $0$ means that no product is recommended. The algorithm is commonly known to the buyer and sellers.

The algorithm can condition on the buyer's values and prices but not on sellers' costs.
This assumption reflects our focus on buyer-side algorithmic mediation.  
The algorithm has buyer information because it is the buyer's interface: it can use buyer history, product attributes, reviews, and other data to form personalized estimates of product values. 
For our baseline analysis, we can treat these estimates directly as the buyer's values $(\wtp_1,\ldots,\wtp_J)$. 
At the same time, sellers' costs remain private because they may depend on procurement conditions, inventory, or other seller-side information that is not observable from the buyer's interface.

\paragraph{Timing.}
At the outset, a designer commits to an algorithm. Then, the game unfolds as follows. First, nature draws seller costs and buyer values. Second, each seller privately observes her own cost and simultaneously posts a price. Third, the algorithm observes the value profile and the price profile and recommends one seller or recommends no purchase. If no product is recommended, trade does not occur. If seller $j\in\Jset$ is recommended, the buyer observes the recommendation and price $p_j$ and decides whether to purchase. If product $j$ is purchased, the buyer and seller $j$ obtain ex post payoffs $\wtp_j-p_j$ and $p_j-\type_j$, respectively; otherwise, all players obtain zero. Given an algorithm, the solution concept is perfect Bayesian equilibrium.

This timing assumes that the recommendation rule is chosen before sellers set their prices. 
It is intended to capture prominent buyer assistants or ranking systems whose rules are already in place when sellers choose or update their prices. 
We further assume that the algorithm can act as a gatekeeper, i.e., the buyer can purchase only the product recommended by the algorithm.
This assumption is relevant in online settings where recommendation systems are used to facilitate product discovery and bring products to buyers' attention.\footnote{See, for example, \cite{dinerstein2018consumer} and \cite{lee2023entry}.}

\paragraph{Objective.}
We denote the buyer's ex ante expected payoff by $\BS$, the sum of sellers' expected profits (or seller surplus) by $\SSurplus$, and total surplus by $\TS=\BS+\SSurplus$. We are interested in the welfare possibilities of algorithmic recommendations. In the main part of the paper, we focus on the buyer-aligned part of the Pareto frontier and let the designer maximize
\begin{equation}\label{eq:main-objective}
\Wobj_\weight \triangleq \weight \BS +(1-\weight)\TS,
\end{equation}
where $\weight\in[0,1]$. The case $\weight=0$ is total-surplus maximization, while $\weight=1$ is the buyer-optimal benchmark. Equivalently, because $\Wobj_\weight=\BS+(1-\weight)\SSurplus$, this class places weakly more weight on buyer surplus than on seller surplus. We refer to \eqref{eq:main-objective} as the class of buyer-aligned objectives.\footnote{The same class of objectives is studied, for example, by \cite{baron1982regulating}, \cite{armstrong2007regulation}, \cite{loertscher2019merger,loertscher2019intermediate}.}

This class is natural for AI assistants and other consumer-facing recommendation tools whose value depends on subscription revenue and participation. 
Also, we show that in the single-seller case, the total-surplus-maximizing algorithm also maximizes seller surplus, so buyer-aligned objectives span the entire Pareto frontier of buyer and seller surplus. 
With multiple sellers, objectives that place more weight on seller surplus than on buyer surplus lead to additional complications, which we discuss in \autoref{sectionSellerSurplusObjective}.

For each $\weight\in[0,1]$ and each seller $j\in\Jset$, define the $\weight$-virtual cost
\begin{equation}\label{eq:alpha-virtual-general}
\virtual_{\weight,j}(c_j)\triangleq c_j+\weight\frac{\typemeas_j(c_j)}{\typedens_j(c_j)}.
\end{equation}
For the dummy seller, we adopt the convention that $\virtual_{\weight,0}(\type_0)=0$. Throughout the paper, we impose the following regularity condition.\footnote{
This is the standard condition that avoids ironing. For nonmonotone (weighted) virtual costs, we can use their ironed versions to prove an analogue of \autoref{prop:competing-sellers} in which the maximized value of the objective is the same under the optimal mechanism and the optimal algorithm. \autoref{prop:single-weight-optimal} holds once we use the generalized inverse of the ironed virtual cost.
}
\begin{assumption}[Regular Virtual Costs]\label{ass:regular-virtual-costs}
For every seller $j\in\Jset$, $\virtual_{1,j}(c_j)$ is continuous and strictly increasing on $[0,1]$.
\end{assumption}

Given an algorithm and an equilibrium, a price is \emph{active} if it results in a strictly positive trade probability and \emph{inactive} otherwise. A type is \emph{active} if she posts an active price with a strictly positive probability and \emph{inactive} otherwise.

\section{Analysis}\label{sec:analysis}

\subsection{Known Seller Costs}
\label{sec:known_cost}

To illustrate the power of algorithmic recommendations, in this section we depart from the baseline model in \autoref{sec:model} by assuming that sellers' costs are commonly known.
The algorithm can then implement the efficient allocation while leaving all surplus to the buyer.
\begin{claim}\label{claim:known_cost}
Suppose that each seller $j$ has a commonly known cost.
There exists an algorithm under which an equilibrium implements the efficient allocation and gives every seller zero profit.
For every weight $\weight\in[0,1]$, the resulting outcome maximizes the objective in \eqref{eq:main-objective} across all algorithms and equilibria.
\end{claim}
\noindent\emph{Proof.} See Appendix~\ref{app:proof-known-cost}.\hspace*{\fill}$\square$\par\smallskip

One such algorithm recommends seller $j\in\Jset$ such that $p_j=c_j$ and
\begin{equation*}
j\in\argmax_{k\in\Jzero}\{\wtp_k-c_k\},
\end{equation*}
and recommends no purchase if such a seller does not exist.
In equilibrium, all sellers set a price equal to their marginal costs, and the buyer purchases from a seller whose value net of cost is maximal and positive.

\autoref{claim:known_cost} illustrates the power of algorithmic recommendations.
However, the optimal algorithm relies on the known-cost assumption. 
In many applications, sellers' costs reflect conditions that are unobservable to the algorithm. 
Thus, in the next section, we turn to our main setup, in which sellers are privately informed about their costs.

\subsection{General Case}\label{subsec:implementation}

The algorithm controls information about the value and existence of sellers but does not administer monetary transfers or specify production allocation directly. 
This, a priori, makes algorithms less powerful than standard mechanisms. 
Nonetheless, we show that algorithms are as powerful as mechanisms in maximizing objectives given in \eqref{eq:main-objective}.

A \emph{direct mechanism} specifies ex post allocation probabilities $q_j(\vecwtp,\vectype)$ and monetary transfers $t_j(\vecwtp,\vectype)$ from the buyer to each seller $j\in\Jset$ as functions of value and cost profiles.
We say that two ex post product allocations, whether induced by direct mechanisms or by algorithms and their equilibria, are the same if they agree for each seller and for almost every $(\vecwtp,\vectype)$ under the joint distribution of values and costs.
The same convention applies to other objects such as a seller's interim-profit schedule. 

Given a direct mechanism, define interim outcomes for each seller $j$:
\begin{equation}\label{eq:interim-quantities}
\begin{aligned}
    Q_j(c_j) &\triangleq \mathbb{E}_{\vecwtp, \vectype_{-j}}[q_j(\vecwtp, c_j, \vectype_{-j})],\\
    T_j(c_j) &\triangleq \mathbb{E}_{\vecwtp, \vectype_{-j}}[t_j(\vecwtp, c_j, \vectype_{-j})],\\
    V_j(c_j) &\triangleq \mathbb{E}_{\vecwtp, \vectype_{-j}}[v_j\, q_j(\vecwtp, c_j, \vectype_{-j})].
\end{aligned}
\end{equation}
Here, $Q_j$ is the interim allocation, $T_j$ is the interim expected revenue, and $V_j$ is the expected value delivered to the buyer from seller $j$ of type $c_j$. As is standard, a mechanism is \emph{incentive compatible} if, for every $j\in\Jset$ and $c_j,c'_j\in[0,1]$, $T_j(c_j)-c_jQ_j(c_j)\ge T_j(c'_j)-c_jQ_j(c'_j)$. A mechanism is \emph{seller individually rational} if, for every $j\in\Jset$ and $c_j\in[0,1]$, $T_j(c_j)-c_jQ_j(c_j)\ge0$.

\begin{lemma}[Implementation Lemma]\label{lemmaImplementation}
Take any direct mechanism whose interim outcomes $(Q_j,T_j, V_j)_{j\in\Jset}$ satisfy the following conditions: for each seller $j$,
		\begin{enumerate}
		    \item $Q_j(c_j)$ is nonincreasing in $c_j$,
            \item $T_j(c_j)=c_jQ_j(c_j)+\int_{c_j}^{1}Q_j(x)\,\dint x$ for all $c_j\in[0,1]$,
		    \item \label{cond:implementation-buyer-obedience}$V_j(c_j)-T_j(c_j)\geq0$ for all $c_j\in[0,1]$.
		\end{enumerate}
Then, an algorithm and an equilibrium exist that attain the same buyer surplus, total surplus, and each seller's interim profits.
Conversely, if interim outcomes $(Q_j,T_j, V_j)_{j\in\Jset}$ are induced by some algorithm and equilibrium, they satisfy the conditions above.
\end{lemma}
\noindent\emph{Proof.} See Appendix~\ref{app:proof-implementation-lemma}.\hspace*{\fill}$\square$\par\smallskip

Conditions 1 and 2 follow from the incentive-compatibility constraints when each seller with the highest cost type earns zero profit.
Condition 3 is imposed by the fact that the algorithm cannot force the buyer to purchase. 
The condition is stronger than ex ante buyer individual rationality because it requires nonnegative buyer surplus from each seller conditional on the seller's cost type.

The construction of the algorithm in \autoref{lemmaImplementation} converts any direct mechanism that satisfies Conditions~1--3 into an algorithm.
Given interim outcomes $(Q_j,T_j,V_j)_{j\in\Jset}$, each seller type $c_j$ is assigned the equilibrium price $p_j^*(c_j)=T_j(c_j)/Q_j(c_j)$ whenever $Q_j(c_j)>0$.
The algorithm observes these prices, backs out the cost levels consistent with those prices, and then reproduces the allocation rule that attains the same interim outcomes.
The algorithm can do so even though the price map $p^*_j(c_j)$ may not be invertible, because the interim allocation rule is measurable with respect to the equilibrium prices.
Conditions 1 and 2 ensure the pricing rule is consistent with seller equilibrium, and Condition 3 ensures buyer obedience.

We use \autoref{lemmaImplementation} to establish the equivalence between optimal mechanisms and optimal algorithms.
\begin{definition}[$\weight$-Optimal Mechanism]\label{def:alpha-optimal}
Fix $\weight\in[0,1]$. An \emph{$\weight$-optimal mechanism} is a direct mechanism that maximizes objective $\Wobj_\weight$  among all incentive-compatible and seller individually rational direct mechanisms.  For $\weight=0$, we impose the normalization that the highest-cost type of each seller earns zero interim profit.
\end{definition}
\begin{definition}[$\weight$-Optimal Algorithm]\label{def:alpha-optimal-algo}
Fix $\weight\in[0,1]$. 
An \emph{$\weight$-optimal algorithm} is a recommendation algorithm under which an equilibrium exists that maximizes objective $\Wobj_\weight$ across all algorithms and equilibria.
\end{definition}

Hereafter, by an $\weight$-optimal mechanism or algorithm, we mean a mechanism or algorithm together with an associated equilibrium that maximizes $\Wobj_\weight$. 

\begin{theorem}\label{prop:competing-sellers}\label{propAlgorithmMechanismEquivalence}
For every $\weight\in[0,1]$, any $\weight$-optimal mechanism and any $\weight$-optimal algorithm attain the same buyer surplus and total surplus, and thus the same value of $\Wobj_\weight$.
Also, they have the same ex post product allocation.

\end{theorem}
\noindent\emph{Proof.} See Appendix~\ref{app:proof-competing-sellers}.\hspace*{\fill}$\square$\par\smallskip

The key observation behind \autoref{prop:competing-sellers} is that the interim outcomes induced by the optimal mechanism, which does not impose any buyer individual rationality constraint, satisfy Condition 3 of \autoref{lemmaImplementation}.
The intuition is as follows.
Under the optimal mechanism, which maximizes weighted virtual surplus, seller $j$'s information rent is determined by how long she would keep the allocation as her reported cost rises, and she loses it at the latest once her cost exceeds the buyer's value $v_j$.
Thus, the rent is bounded by the buyer's value net of the actual cost, so the incentive-compatible transfer never exceeds the expected buyer value generated by the allocation, which is Condition 3.\footnote{Formally, fix seller $j$, type $c_j$, and a realization $(\vecwtp,\vectype_{-j})$, and let $r_j(x)=q_j(\vecwtp,x,\vectype_{-j})$. The allocation rule implies that $r_j(x)=0$ for $x>v_j$ and that $r_j(x)$ is weakly decreasing in $x$. Therefore $\int_{c_j}^1 r_j(x)\,\dint x\le (v_j-c_j)r_j(c_j)$. Taking expectations gives $T_j(c_j)=c_jQ_j(c_j)+\int_{c_j}^1Q_j(x)\,\dint x\le V_j(c_j)$.}

We next describe an optimal algorithm constructed in the proof of \autoref{prop:competing-sellers}.
Fix any $\weight\in[0,1]$.
For each seller $j\in\Jset$, define the auxiliary random variable
\begin{equation}\label{eq:theta-alpha}
\theta_{\weight,j}=\wtp_j-\max_{k\in\Jzero\setminus\{j\}}\{\wtp_k-\virtual_{\weight,k}(\type_k)\},
\end{equation}
that is, seller $j$'s value net of the highest $\weight$-virtual surplus among all other sellers, including the no-purchase option.

For each seller $j\in\Jset$, let $\win_{\weight,j}$ denote the set of types $c_j$ such that $\theta_{\weight,j}>\virtual_{\weight,j}(c_j)$ could arise with strictly positive probability.
On realizations relevant for trade, $\virtual_{\weight,j}^{-1}(\theta_{\weight,j})$ has a simple interpretation as seller $j$'s critical cost: holding values and rivals' costs fixed, it is the highest cost at which seller $j$ remains eligible for selection under the weighted-virtual-surplus
rule. Thus, up to tie-breaking, type $c_j$ trades if and only if this critical cost is at least $c_j$.
For every type $c_j\in\win_{\weight,j}$, define the price
\begin{equation}\label{eq:general-price-alpha}
 p^*_{\weight,j}(c_j)\triangleq \E_{\theta_{\weight,j}}\!
 \left[\virtual_{\weight,j}^{-1}(\theta_{\weight,j})
 \mid \theta_{\weight,j}\ge \virtual_{\weight,j}(c_j)\right],
\end{equation}
and extend the domain of $p^*_{\weight,j}$ to include types outside of $\win_{\weight,j}$ by assigning some price outside of $p^*_{\weight,j}(\win_{\weight,j})$.
The price is the expected critical cost conditional on trade. Finally, let $C_j(p_j)$ denote the set of types that post $p_j$ under \eqref{eq:general-price-alpha}, i.e., 
$$
C_j(p_j)\triangleq\{c_j\in[0,1]\mid p^*_{\weight,j}(c_j)=p_j\}.
$$
In the constructed algorithm and its associated equilibrium, each seller $j\in\Jset$ posts $p^*_{\weight,j}(c_j)$. For almost every profile $(\vecwtp,\vecp)$ of values and on-path prices, there exists a seller $j^*(\vecwtp,\vecp)\in\Jzero$ such that
\begin{equation}\label{equationJ}
j^*(\vecwtp,\vecp)\in\argmax_{j\in\Jzero}
\{\wtp_j-\virtual_{\weight,j}(c_j)\}
\end{equation}
for every type profile $\vectype$ with $c_j\in C_j(p_j)$ for all $j\in\Jset$. The algorithm recommends seller $j^*(\vecwtp,\vecp)$.

Intuitively, the equilibrium prices encode the seller cost information the algorithm needs. 
Given prices, the algorithm recommends a seller with the highest nonnegative weighted virtual surplus among the types consistent with those prices. 
Higher weight $\weight$ raises the information-rent term in \eqref{eq:alpha-virtual-general}, which makes the algorithm more willing to exclude high-cost sellers in order to improve the buyer's payoff. 

Notably, under the optimal algorithm, the product value distribution $\wtpmeas$ affects readily observable variables, the algorithmic demand and equilibrium prices, but not the product allocation rule: the product is always allocated to a seller with the highest nonnegative weighted virtual surplus. This feature is crucial for the market-segmentation neutrality results in \autoref{sectionData}.

\section{Single Seller}\label{sec:single}

In this section, we consider the case of a single seller, which allows us to write the optimal algorithm in a simpler form.
It also permits stronger welfare analysis: 
in \autoref{subsec:payoff-set}, we characterize the entire set of buyer and seller surplus pairs attainable across all algorithms. 
For the rest of the section, we suppress seller subscripts.

\subsection{Optimal Algorithm}\label{subsec:single-optimal}

We define a statistic that converts the desired weighted-virtual-cost allocation into a price threshold. For each $v\in[0,1]$, define the $\weight$-\strategic value

\begin{equation}\label{equationStrategicAlphaBS}
\vst_\weight(v)\triangleq \expect_{\V\sim G}\left[\virtual_\weight^{-1}(\V)\mid \V\ge v\right].
\end{equation}
Whenever $\Pr[\V\ge v]=0$, set $\vst_\weight(v)=\virtual_\weight^{-1}(v)$; this convention affects only $G$-null values. The \strategic value $\vst_\weight(v)$ is weakly increasing in $v$. When $v$ is sufficiently low, $\vst_\weight(v)>v$ because $\vst_\weight(0)=\expect_{\V\sim G}[\virtual_\weight^{-1}(\V)]>0$. If $\weight>0$, then $\vst_\weight(1)=\virtual_\weight^{-1}(1)<1$, so $\vst_\weight(v)<v$ for sufficiently high $v$.
Let $\overline c_\weight\triangleq\virtual_\weight^{-1}(1)$.

\begin{corollary}
[A Single Seller]\label{prop:single-weight-optimal}
An $\weight$-optimal algorithm recommends the product if and only if $\vst_\weight(v)\ge p$. 
In the equilibrium, the seller of type $c\le\overline c_\weight$ posts price $p^*_\weight(c)=\vst_\weight(\virtual_\weight(c))$, and the seller of type $c>\overline c_\weight$ is inactive. 
Trade occurs if and only if $v\ge\virtual_\weight(c)$.
\end{corollary}
\noindent\emph{Proof.} See Appendix~\ref{app:proof-single-weight-optimal}.\hspace*{\fill}$\square$\par\smallskip

Take $\weight>0$, and compare the $\weight$-optimal algorithm with the myopic algorithm, which recommends if and only if $\wtp \ge p$.
Then, the optimal algorithm overrecommends at low prices and underrecommends at high prices.
For seller types that satisfy $\vst_\weight(\wtp) < \vst_\weight(\virtual_\weight(\type)) < \wtp$, the \algo does not recommend the product even though its value exceeds the price, whereas for seller types that satisfy $\wtp < \vst_\weight(\virtual_\weight(\type)) < \vst_\weight(\wtp)$, the \algo recommends the product even though its value is below the price. 
Such a policy, which punishes high prices and rewards low prices compared to the myopic rule, exerts downward pressure on prices.

Under the total-surplus-maximizing algorithm (i.e., $\weight=0$), the underrecommendation force disappears, while the overrecommendation force remains. 
When $\wtpmeas$ is continuous with full support on $[0,1]$, the threshold representation is equivalent to recommending the product for every $\wtp$ if $p<\expect[\wtp]$, and, otherwise,  recommending it if and only if $\wtp\ge\wtptr(p)$, where
\begin{equation*}
\expect[v\mid v\ge\wtptr(p)]=p.
\end{equation*}
Under this \algo, the buyer is indifferent between purchase and no purchase at every price $p\ge \expect[\wtp]$, while every price $p<\expect[\wtp]$ is dominated for the seller by $p=\expect[\wtp]$.
Thus, the seller captures all gains from trade, so profit maximization aligns with efficiency. A seller of type $\type$ posts $\popt_0(\type)=\expect[\wtp\mid\wtp\ge \type]$, which induces trade if and only if $\wtp\ge\type$. The resulting allocation is efficient, the seller obtains the maximal feasible seller surplus (hence the \algo is also seller-optimal), and the buyer surplus is zero.\footnote{\citet{gottardi2024shuttle} use a similar argument to construct an efficient one-shot mediation mechanism.}

The uniform case makes the optimal algorithm, the equilibrium prices, and both distortions explicit.
\begin{example}[Uniform]\label{exampleUniform}
Let $\type$ and $\wtp$ be uniformly distributed on $[0,1]$.
In this case, the $\weight$-virtual cost is $\virtual_\weight(\type)=(1+\weight)\type$, so the active-cost cutoff is $\overline c_\weight=1/(1+\weight)$. The $\weight$-\strategic value is $\vst_\weight(\wtp)= \E_{\tilde \wtp \sim U[0,1]}\left[\tilde \wtp/(1+\weight)\mid \tilde \wtp \ge \wtp\right]=(1+\wtp)/(2(1+\weight))$.
Because this \strategic value is strictly increasing, the $\alpha$-optimal recommendation rule can be written as a value cutoff: for prices in the active range, the algorithm recommends the product if and only if $\wtp\ge\wtptr_\weight(p)$, where $\vst_\weight(\wtptr_\weight(p))=p$. Thus $\wtptr_\weight(p)$ is equal to $0$ for $p\le 1/(2(1+\weight))$, to $2(1+\weight)p-1$ for $p\in[1/(2(1+\weight)),1/(1+\weight)]$, and to $1$ for $p\ge 1/(1+\weight)$; the cutoff values $0$ and $1$ mean, respectively, that the product is recommended for every realization of $\wtp$ and only when $\wtp=1$, a zero-probability event.

The equilibrium price posted by active type $\type$ is $\popt_\weight(\type)=\vst_\weight(\virtual_\weight(\type))=1/(2(1+\weight))+\type/2$ for $\type\in[0,1/(1+\weight)]$.\footnote{Types $\type>1/(1+\weight)$ are inactive and post, for example, the limiting active price $1/(1+\weight)$.}
A buyer who receives a recommendation to purchase at price $p\in[1/(2(1+\weight)),1/(1+\weight)]$ infers that the product's expected value is $(1+\weight)p$, and is thus willing to purchase it.

The left side of \autoref{fig:algorithm} depicts the optimal recommendation threshold for $\weight=1$ and $\weight=1/2$, along with the myopic recommendation threshold (dashed line). Increasing $\weight$ raises the recommendation threshold at every interior price and shifts the relevant price cutoffs to the left, so high prices are punished more aggressively.
For any $\weight>0$, the $\weight$-optimal \algo is suboptimal ex post in two ways: if the product price is low, i.e., $p<1/(1+2\weight)$, it recommends the product even when the value is below the price; if the product price is high, i.e., $p>1/(1+2\weight)$, the \algo does not recommend the product even when the value is above the price.

\begin{figure}
\centering
\begin{tikzpicture}
        \begin{axis}[
        width=0.45\textwidth,
        height=0.45\textwidth,
            xmin=0, xmax=1.1,
            ymin=0, ymax=1.1,
            axis lines=middle,
            xtick={0.5,1},
            ytick={0.5,1},
            xlabel={$p$},
            ylabel={$v$},
            xlabel style={at={(ticklabel* cs:1)}, anchor=north west, shift={(0ex,0ex)}},
            ylabel style={at={(ticklabel* cs:1)}, anchor=south east, shift={(0ex,0ex)}},
        ]
            \addplot[dashed,very thin,black!65] coordinates {(0,0) (1,1)};
            \addplot[thick,myblue] coordinates {(0,0) (1/4,0) (1/2,1) (1,1)};
            \addplot[thick,mygreen] coordinates {(0,0) (1/3,0) (2/3,1) (1,1)};
            \node[black!65] at (axis cs:0.88,0.70) {$v=p$};
            \node[myblue] at (axis cs:0.27,0.81) {$\weight=1$};
            \node[mygreen] at (axis cs:0.7,0.4) {$\weight=1/2$};
        \end{axis}
\end{tikzpicture}
\hspace{0.5cm}
\begin{tikzpicture}
        \begin{axis}[
        width=0.45\textwidth,
        height=0.45\textwidth,
            xmin=0, xmax=1.1,
            ymin=0, ymax=1.1,
            axis lines=middle,
            xtick={0.5,1},
            ytick={0.5,1},
            xlabel={$c$},
            ylabel={$v$},
            xlabel style={at={(ticklabel* cs:1)}, anchor=north west, shift={(0ex,0ex)}},
            ylabel style={at={(ticklabel* cs:1)}, anchor=south east, shift={(0ex,0ex)}},
        ]
            \fill[blue!20] (axis cs:0,0) -- (axis cs:1/2,1) -- (axis cs:0,1) -- cycle;
            \addplot[thick] coordinates {(0,1/4) (1/2,1/2)(1,1/2)};
            \addplot[thick] coordinates {(0,0) (1/2,1)};
            \addplot[dashed,very thin,black!65] coordinates {(0,0) (1,1) (0,1) (0,0)};
            \node at (axis cs:0.65,0.86) {$\virtual_1(c)=2c$};
            \draw[->, thick, >=stealth] (axis cs:0.54,0.81) -- (axis cs:0.4,0.75);
            \node at (axis cs:0.66,0.37) {$\popt_1(c)$};
            \draw[->, thick, >=stealth] (axis cs:0.55,0.38) -- (axis cs:0.31,0.39);
        \end{axis}
\end{tikzpicture}
    \caption{Optimal recommendation algorithm (left) and the resulting equilibrium pricing strategy and trade region for $\weight=1$ (right). $J=1$, $\wtp\sim U[0,1]$, $\type\sim U[0,1]$.}
    \label{fig:algorithm}
\end{figure}

The right side of \autoref{fig:algorithm} depicts, for the buyer-optimal case $\weight=1$, the price $\popt_1(\type)$ posted by the seller of type \type, the region of values and types in which trade occurs (filled area), and the efficient trade region (area encircled by dashed lines).
By \autoref{prop:single-weight-optimal}, trade occurs whenever the buyer value exceeds the seller's $\weight$-virtual cost. Type $\type=0$ always trades; all higher active types post progressively higher prices and serve progressively fewer buyers. Types $\type>1/(1+\weight)$ never trade. Equilibrium active prices span the interval $[1/(2(1+\weight)),1/(1+\weight)]$.

We now compare the equilibrium outcomes under the $\weight$-optimal algorithm and the myopic algorithm.
Under the myopic algorithm, the seller's problem is a standard monopoly problem with the optimal price $p^m(\type) =  (1+\type)/2$, and trade occurs if and only if $\wtp \ge p^m(\type)$.

For every active type, $p^m(\type)-\popt_\weight(\type)=\weight/(2(1+\weight))$. Thus, increasing $\weight$ shifts the seller's pricing function downward in parallel for all types that remain active, while the active set shrinks from $[0,1)$ at $\weight=0$ to $[0,1/2)$ at $\weight=1$, as in the left panel of \autoref{fig:algorithm2}.

\begin{figure}
\centering
\begin{tikzpicture}
    \begin{axis}[
        width=0.45\textwidth,
        height=0.45\textwidth,
        xmin=0, xmax=1.1,
        ymin=0, ymax=1.1,
        axis lines=middle,
        xtick={0.5,1},
        ytick={0.25, 0.5,1},
        xlabel={$\type$},
        ylabel={$p$},
        xlabel style={at={(ticklabel* cs:1)}, anchor=north west, shift={(0ex,0ex)}},
        ylabel style={at={(ticklabel* cs:1)}, anchor=south east, shift={(0ex,0ex)}},
    ]
        \addplot[thick,black!70] coordinates {(0,0.5) (1,1)};
        \addplot[thick,myblue] coordinates {(0,0.25) (0.5,0.5)};
        \addplot[thick,mygreen] coordinates {(0,1/3) (2/3,2/3)};
        \addplot[dotted, thin,myblue] coordinates {(0,0.5) (0.5,0.5) (0.5,0)};
        \addplot[dotted, thin,mygreen] coordinates {(0,2/3) (2/3,2/3) (2/3,0)};
        \node[black!70] at (axis cs:0.76,0.58) {$p^m(c)$};
        \draw[->, thick, >=stealth] (axis cs:0.75,0.64) -- (axis cs:0.7,0.8);
        \node[myblue] at (axis cs:0.35,0.17) {$\popt_1(c)$};
        \draw[->, thick, >=stealth, myblue] (axis cs:0.29,0.20) -- (axis cs:0.16,0.30);
        \node[mygreen] at (axis cs:0.67,0.37) {$\popt_{1/2}(c)$};
        \draw[->, thick, >=stealth, mygreen] (axis cs:0.54,0.42) -- (axis cs:0.35,0.49);
    \end{axis}
\end{tikzpicture}
\hspace{0.5cm}
\begin{tikzpicture}
    \begin{axis}[
        width=0.45\textwidth,
        height=0.45\textwidth,
        xmin=0, xmax=1.1,
        ymin=0, ymax=1.1,
        axis lines=middle,
        xtick={0.5,1},
        ytick={0.5,1},
        xlabel={$c$},
        ylabel={$v$},
        xlabel style={at={(ticklabel* cs:1)}, anchor=north west, shift={(0ex,0ex)}},
        ylabel style={at={(ticklabel* cs:1)}, anchor=south east, shift={(0ex,0ex)}},
    ]
        \fill[red!30] (axis cs:1/3,2/3) -- (axis cs:1/2,1) -- (axis cs:1,1) -- cycle;
        \fill[blue!30] (axis cs:1/3,2/3) -- (axis cs:0,0) -- (axis cs:0,0.5) -- cycle;
        \fill[black!20] (axis cs:0,0.5)--(axis cs:1/3,2/3) -- (axis cs:0.5,1) -- (axis cs:0,1) -- cycle;
        \addplot[thick] coordinates {(0,0) (1/2,1)};
        \addplot[thick] coordinates {(0,0.5) (1,1)};
        \node at (axis cs:0.46,0.17) {$\virtual_1(c)=2c$};
        \draw[->, thick, >=stealth] (axis cs:0.35,0.2) -- (axis cs:0.18,0.3);
        \node at (axis cs:0.79,0.58) {$p^m(c)$};
        \draw[->, thick, >=stealth] (axis cs:0.75,0.64) -- (axis cs:0.7,0.8);
    \end{axis}
\end{tikzpicture}
    \caption{Comparison of equilibrium outcomes under the $\weight$-optimal algorithm and the myopic algorithm in terms of the seller's pricing strategies (left) and trade regions for $\weight=1$ (right). $J=1$, $\wtp\sim U[0,1]$, $\type\sim U[0,1]$.}
    \label{fig:algorithm2}
\end{figure}

The right panel of \autoref{fig:algorithm2} depicts the sets of pairs of value $\wtp$ and cost $\type$ under which trade occurs for $\weight=1$.
Compared with the myopic algorithm, the buyer-optimal algorithm reduces trade between high-value buyers and high-cost sellers (the red region) and increases trade between low-value buyers and low-cost sellers (the blue region). We generalize this observation in \autoref{subsec:allocation-substitution}.
\hfill$\closing$
\end{example}

\subsection{The Payoff Set}\label{subsec:payoff-set}
Focusing on the case of a single seller, we now answer a broader question: beyond optimal algorithms, what is the range of welfare outcomes that algorithmic recommendations can generate? Formally, we characterize the \emph{payoff set}
\begin{equation*}
\payoffset\triangleq\left\{(\BS,\SSurplus)\;\middle|\;\text{some algorithm and equilibrium attain $(\BS,\SSurplus)$}\right\}.
\end{equation*}

The characterization relies on the family of algorithms in \autoref{prop:single-weight-optimal}.
For any $\weight\ge0$, including $\weight> 1$, the $\weight$-virtual cost \eqref{eq:alpha-virtual-general} and the $\weight$-\strategic value \eqref{equationStrategicAlphaBS} remain well defined, provided that $\virtual_\weight$ is strictly increasing. 
To guarantee this property for every $\weight\ge0$, in this section, we assume that $\typemeas/\typedens$ is nondecreasing on $[0,1]$.

We refer to the algorithm and equilibrium of \autoref{prop:single-weight-optimal} with parameter $\weight\ge0$ as the \emph{$\weight$-algorithm} and denote the resulting payoffs by $(\BS_\weight,\SSurplus_\weight)$; the construction and the equilibrium verification apply verbatim to any $\weight\ge0$.
Under the $\weight$-algorithm, trade occurs if and only if $\wtp\ge\virtual_\weight(\type)$.
For $\weight\in[0,1]$, the $\weight$-algorithm is $\weight$-optimal. For $\weight>1$, it maximizes $\BS-(\weight-1)\SSurplus$, an objective that penalizes seller surplus.

\begin{proposition}
	\label{prop:payoff-set}
	With a single seller and increasing $\typemeas/\typedens$, 
\begin{equation}
\payoffset=\mathrm{conv}\left(\{(0,0)\}\cup\left\{(\BS_\weight,\SSurplus_\weight)\mid\weight\ge0\right\}\right),
\end{equation}
and $\left\{(\BS_\weight,\SSurplus_\weight)\mid\weight\in[0,1]\right\}$ is the Pareto frontier of $\payoffset$.\footnote{$\mathrm{conv}(A)$ denotes the convex hull of a set $A$. A point of $\payoffset$ is on the Pareto frontier if no other point of $\payoffset$ weakly increases both buyer surplus and seller profit, with at least one strict increase.}
Moreover, $\payoffset$ coincides with the payoff set of direct mechanisms that are incentive compatible, seller individually rational, and attain a nonnegative buyer surplus.
\end{proposition}
\noindent\emph{Proof.} See Appendix~\ref{app:proof-payoff-set}.\hspace*{\fill}$\square$\par\smallskip

\begin{figure}[!t]
\centering
\begin{tikzpicture}
    \begin{axis}[
        width=0.62\textwidth,
        height=0.55\textwidth,
        xmin=0, xmax=0.12,
        ymin=0, ymax=0.185,
        axis lines=middle,
        xtick={0.0416667,0.0833333},
        xticklabels={$\tfrac{1}{24}$,$\tfrac{1}{12}$},
        ytick={0.0416667,0.0833333,0.1666667},
        yticklabels={$\tfrac{1}{24}$,$\tfrac{1}{12}$,$\tfrac{1}{6}$},
        xlabel={$\BS$},
        ylabel={$\SSurplus$},
        xlabel style={
            at={(ticklabel* cs:1)},
            anchor=north west,
            shift={(0ex,0ex)}
        },
        ylabel style={
            at={(ticklabel* cs:1)},
            anchor=south east,
            shift={(0ex,0ex)}
        },
        clip=false,
    ]
        \addplot[
            fill=blue!15,
            draw=none
        ]
        plot[
            domain=0:0.1666667,
            samples=150
        ]
        ({sqrt(2*x/3)-2*x},{x}) -- cycle;

        \addplot[
            very thick,
            myblue,
            domain=0.0416667:0.1666667,
            samples=100
        ]
        ({sqrt(2*x/3)-2*x},{x});

        \addplot[
            thick,
            mygreen,
            domain=0:0.0416667,
            samples=100
        ]
        ({sqrt(2*x/3)-2*x},{x});

        \addplot[
            only marks,
            mark=*,
            mark size=1.6pt,
            black
        ]
        coordinates {
            (0,0.1666667)
            (0.0833333,0.0416667)
            (0,0)
            (0.0416667,0.0833333)
        };

        \node[
            anchor=north east,
            font=\small
        ]
        at (axis cs:0,0)
        {$(0,0)$};

        \node[
            anchor=west,
            font=\small
        ]
        at (axis cs:0.004,0.1700)
        {seller optimum = efficient, $\weight=0$};

        \node[
            anchor=west,
            font=\small
        ]
        at (axis cs:0.0865,0.0416667)
        {buyer optimum, $\weight=1$};

        \node[
            anchor=west,
            font=\small
        ]
        at (axis cs:0.004,0.008)
        {no trade};

        \node[
            anchor=east,
            align=center,
            font=\small
        ]
        at (axis cs:0.063,0.065)
        {myopic\\algorithm};

        \node[
            myblue,
            anchor=south west,
            font=\small
        ]
        at (axis cs:0.054,0.111)
        {Pareto frontier};

        \node[
            mygreen,
            anchor=west,
            font=\small
        ]
        at (axis cs:0.069,0.011)
        {$\alpha>1$};
    \end{axis}
\end{tikzpicture}
    \caption{The payoff set. $J=1$, $\wtp\sim U[0,1]$, $\type\sim U[0,1]$.}
    \label{fig:payoff-set}
\end{figure}

The result relies on the fact that the efficient algorithm, i.e., $\weight =0$, which attains zero buyer surplus, is also seller-optimal.
As a result, the efficient algorithm maximizes every objective that places weakly more weight on seller profit than on buyer surplus. 
The complementary objectives, which place weakly more weight on buyer surplus, are given by \eqref{eq:main-objective}.
Thus, the algorithms characterized in \autoref{prop:single-weight-optimal} span the Pareto frontier.
The outcomes induced by the $\alpha$-algorithms with $\alpha>1$, along with the vertical segment that connects the no-trade point $(0,0)$ to the seller-optimal point, span the rest of the boundary. 
The $\alpha>1$ curve is Pareto dominated by the buyer-optimal point, while the vertical segment is dominated by the seller-optimal point.

\autoref{fig:payoff-set} depicts the payoff set under the uniform value and cost distributions.\footnote{See Appendix~\ref{app:uniform-payoff-set} for the derivation.}
As illustrated there, the payoff set generally has the following properties.
Along the boundary, as the weight $\weight$ on buyer surplus increases from zero, the payoff point moves away from the efficient (and thus seller-optimal) point and buyer surplus rises.
Simultaneously, the trade event $\{(\type,\wtp)\mid\wtp\ge\virtual_\weight(\type)\}$ shrinks and total surplus decreases.
Once $\weight$ exceeds $1$, the boundary bends back toward the no-trade point $(0,0)$.

The last part of \autoref{prop:payoff-set} extends the equivalence of \autoref{prop:competing-sellers} from optimal outcomes to the entire payoff set.
This equivalence is specific to monopoly: with competing sellers, seller-favored objectives create a wedge between algorithms and mechanisms (see \autoref{sectionSellerSurplusObjective}).

\section{Economic Implications}\label{sec:implications}

\subsection{Algorithmic Recommendations and Seller Competition}\label{subsec:no-algorithm}\label{subsec:competition-effects}

Algorithmic recommendations, via commitment, provide the buyer with bargaining power that ordinary price competition may not. 
To measure their value, we compare algorithmic recommendations with the game of \emph{Bertrand competition}.
In this game, each seller privately observes her cost and simultaneously posts a price.
Then, the buyer, after observing the realized values and prices, purchases the product that yields the highest nonnegative net value.
Under Bertrand competition, the buyer is assumed to be fully informed about products' values and existence.
Because the algorithm is also fully informed about them, this assumption places the two settings on an equal footing.

Bertrand competition is \emph{improvable} for an objective if some algorithm and equilibrium yield a strictly larger value of that objective.
Algorithmic recommendations improve on Bertrand competition for every objective that puts positive weight on buyer surplus:

\begin{proposition}[Improvability over Bertrand Competition]\label{propStrictImprovabilityBSTS}
Suppose values are mutually independent, continuously distributed on $[0,1]$, and have positive densities on $(0,1)$. For every $\weight\in(0,1]$, every equilibrium under Bertrand competition is improvable for the objective $\Wobj_\weight$.
\end{proposition}
\noindent\emph{Proof.} See Appendix~\ref{proofpropStrictImprovabilityBSTS}.\hspace*{\fill}$\square$\par\smallskip
The intuition is as follows.
Under Bertrand competition, every seller whose cost is strictly below the top of the support trades with positive probability.
Indeed, seller $j$ with cost $\type_j<1$ can earn a positive profit by posting a price of $p = (1+\type_j)/2$, which leads to sales if $\wtp_j - p > \wtp_k$.
By contrast, an $\weight$-optimal algorithm with $\weight>0$ excludes a top interval of costs through the weighted virtual-cost cutoff. 
These excluded types may create positive total surplus, but serving them would require prices or information rents that are too high for a buyer-aligned objective.
Algorithmic recommendations thus improve on Bertrand competition by committing the buyer to a price- and value-contingent access rule that ordinary competition cannot implement.

We now turn to the next question: how the gain from the optimal algorithm depends on the degree of competition.
We study this question through two comparative statics: increasing the number of sellers and merging two sellers. 
For tractability, we consider a common-value benchmark in which all sellers have the same deterministic value $v\in(0,1]$ and i.i.d. costs, abstracting from product-differentiation effects and focusing on how competition and algorithmic recommendations discipline privately informed sellers.

For a fixed weight $\weight \in [0,1]$, we compare the value of competition with and without an algorithm.
Competition without an algorithm is Bertrand competition described above.
In the common-value benchmark, the game admits essentially a unique equilibrium, in which each seller adopts the same pricing strategy, which is strictly increasing in the seller's cost type.\footnote{Precisely, Proposition~1 of \cite{athey2004collusion} implies that the equilibrium strategy of each seller with cost below $v$ is the same across all equilibria.
For a cost type above $v$, posting any price above $v$ is consistent with an equilibrium.}
Denote the lowest and second-lowest realized costs by $c_{(1:J)}$ and $c_{(2:J)}$, respectively.
In equilibrium, trade occurs if and only if $c_{(1:J)}\le v$, and a lowest-cost seller transacts at price $\E[\min\{c_{(2:J)},v\}\mid c_{(1:J)}]$.\footnote{By revenue equivalence, expected payoffs are the same as in a second-price procurement mechanism with price cap $v$, in which the lowest-cost seller receives $\min\{c_{(2:J)},v\}$.}

With the optimal algorithm, \autoref{prop:competing-sellers} and the pricing rule \eqref{eq:general-price-alpha} imply that the equilibrium outcome under the $\weight$-optimal algorithm coincides with that under Bertrand competition in which the buyer's value is $\virtual^{-1}(v)$.
In equilibrium, trade occurs if and only if $c_{(1:J)}\le \virtual^{-1}(v)$, and a lowest-cost seller supplies at price $\E[\min\{c_{(2:J)},\virtual^{-1}(v)\}\mid c_{(1:J)}]$.
The optimal algorithm induces a transaction only with a lowest-cost seller but commits not to induce a transaction if $c_{(1:J)} >\virtual^{-1}(v)$.

Let $\Wobj^A_\weight(J)$ and $\Wobj^B_\weight(J)$ denote the value of the objective $\Wobj_\weight$ with $J$ sellers under the $\weight$-optimal algorithm and Bertrand competition, respectively.

\begin{proposition}[Impact of Market Thickness]\label{claim:weighted-ratio-monotonicity}
Suppose all sellers have the same deterministic value $v\in(0,1]$, costs are i.i.d. with common distribution $F$ and density $f$, and $F/f$ is strictly increasing on $[0,1]$. 
For any $\weight\in(0,1]$, the ratio $\Wobj^A_\weight(J)/\Wobj^B_\weight(J)$ is strictly decreasing in $J$ for all $J\ge2$, and converges to $1$ as $J\to\infty$.
\end{proposition}

\noindent\emph{Proof.} See Appendix \ref{secAppendixWeightedRatioMonotonicity}.\hspace*{\fill}$\square$\par\smallskip

The ratio in \autoref{claim:weighted-ratio-monotonicity} measures the relative value of algorithmic recommendations over ordinary competition. The algorithm improves on Bertrand competition by committing the buyer not to trade when the value is above the lowest cost but below the lowest weighted virtual cost. 
As the number of sellers increases, the lowest and second-lowest costs shift toward the bottom of the support, competitive payments approach costs, and the probability and welfare importance of these marginal exclusion cases decline. Thus, the incremental value of algorithmic recommendations decreases with the number of competitors and vanishes in the competitive limit.

An alternative way to change the degree of competition is a merger. As in \cite{loertscher2019merger,loertscher2022incomplete}, we model a merger of two sellers as replacing them with a single seller whose cost equals the minimum of the two original costs. The effect on welfare is \emph{a priori} ambiguous: because the merged seller's cost is stochastically lower than the pre-merger cost of either seller, the merger may strengthen rather than weaken price competition. 
Nevertheless, we show that the welfare effect is unambiguous.

\begin{proposition}[Impact of Merger]\label{propMergerBSTS}
Suppose that all sellers have the same deterministic value $v\in(0,1]$, costs are i.i.d. with common distribution $F$ and density $f$, and $F/f$ is strictly increasing on $[0,1]$. Fix $\weight>0$. Then a merger of two sellers out of $J \ge 2$ weakly lowers the maximal value of $\Wobj_\weight$.
\end{proposition}

\noindent\emph{Proof.} See Appendix \ref{proofpropMergerBSTS}.\hspace*{\fill}$\square$\par\smallskip

The force behind this result is that the buyer-aligned algorithm screens sellers by weighted virtual cost, not physical cost. The first-best total surplus is unaffected by the merger: the lower of the two merging costs remains available to produce, so the efficient allocation is unchanged. But pooling the two cost draws into their minimum raises the merged seller's weighted virtual cost above the lower of the two pre-merger virtual costs, even though its physical cost is lower. 
Thus, the algorithm must leave the merged seller more rent to secure any given allocation probability, so no post-merger algorithm attains a higher value of $\Wobj_\weight$ than the pre-merger optimum. Every objective that puts positive weight on buyer surplus is thus weakly lower after the merger.

Together, the two comparative statics show that algorithmic recommendations are most valuable when ordinary competition is thin, because committed recommendations substitute for missing competitive discipline. 
However, the algorithm does not eliminate the value of competition. 
Greater competition reduces the relative value of algorithmic recommendations over Bertrand competition, while a merger weakly lowers the maximal value that the algorithm can achieve.

\subsection{Allocation Substitution}\label{subsec:allocation-substitution}

Algorithmic recommendations affect the composition of trade, not only the prices at which a fixed set of trades occurs. Indeed, an algorithm treats seller information rents as part of the cost of trade. 
As a result, it may substitute low-cost, lower-value trades for high-cost, higher-value trades.
We formalize this idea below.

We begin with the case of a single seller.
Let $\Trade^\weight$ denote the set of pairs $(c,v) \in [0,1]^2$ such that trade occurs under the $\weight$-optimal algorithm. 
Also, let $\Trade^m$ denote the set of pairs $(c,v)$ such that trade occurs under the myopic algorithm, taking into account the monopoly pricing it induces. 

\begin{proposition}[Allocation Substitution in Monopoly]\label{propositionComparison} 
Fix $\weight>0$. If $\typedens/\typemeas$ is strictly decreasing and  $\wtpdens/(1-\wtpmeas)$ is strictly increasing, then for any $(c,v)\in\Trade^\weight\setminus\Trade^m$ and $(c',v')\in\Trade^m\setminus\Trade^\weight$ such that $(c,v)\neq(c',v')$, $c<c'$ and $v<v'$.
\end{proposition}
\noindent\emph{Proof.} See Appendix~\ref{proofpropositionComparison}.\hspace*{\fill}$\square$\par\smallskip

The hazard-rate assumptions of \autoref{propositionComparison} imply a single-crossing comparison between the algorithmic and monopoly trade thresholds, which yields the sharp result. 
The result captures the idea that the algorithm favors trades that involve lower costs, because they lead to lower information rents.
With multiple sellers, the same substitution pattern appears in parameterized environments with a common (uncertain) value and i.i.d. costs.\footnote{A general comparison between the optimal algorithm and Bertrand competition is difficult because, with heterogeneous product values, the two regimes can select different sellers, so their trade regions need not be represented by thresholds in a common scalar cost statistic.}
The relevant cost statistic is now the lowest cost, $m=\min_{j\in\Jset}\type_j$, because in either regime, the buyer purchases only from a lowest-cost seller.
In this case, we use $\Trade^\weight$ for the set of pairs $(m,\wtp)\in[0,1]^2$ such that trade occurs under the $\weight$-optimal algorithm.
Let $\Trade^B$ denote the trade set in the unique symmetric equilibrium of the Bertrand benchmark.

\begin{proposition}[Allocation Substitution in Oligopoly]\label{propositionComparisonOligopoly}
Let $J\ge2$ sellers have costs drawn i.i.d. from
$\typemeas(\type)=1-(1-\type)^{x}$ on $[0,1]$ with $x\ge1$,\footnote{For $x>1$, the virtual cost diverges to $+\infty$ as $c\to 1$.
But all of our results hold under the weaker condition that, for each seller $j$, $\virtual_{1,j}$ is continuous and strictly increasing on $[0,1)$ and $\lim_{c_j\uparrow 1}\virtual_{1,j}(c_j)>1$, where the limit may be $+\infty$.
In this case, each $\virtual_{\weight,j}(1)$ should be viewed as its left limit, and the relevant inverses are used only on the value support $[0,1]$.} and let the
unit-demand buyer's common value have distribution
$\wtpmeas(\wtp)=1-(1-\wtp)^{y}$ on $[0,1]$ with $y>0$. Let $m=\min_{j\in\Jset}\type_j$ and fix $\weight>0$. Then for any $(m,\wtp)\in\Trade^{\weight}\setminus\Trade^{B}$ and $(m',\wtp')\in\Trade^{B}\setminus\Trade^{\weight}$, $m<m'$ and $\wtp<\wtp'$.
\end{proposition}
\noindent\emph{Proof.} See Appendix~\ref{proofpropositionComparisonOligopoly}.\hspace*{\fill}$\square$\par\smallskip

The parametric assumptions in \autoref{propositionComparisonOligopoly} make the comparison with Bertrand transparent. They isolate the same single-crossing logic as in the monopoly case: algorithmic recommendations expand trade at lower costs and lower values, and contract trade at higher costs and higher values.

The above results show that compared to the no-algorithm benchmark, the trades added by the algorithm are not simply the highest-value trades; they are lower-cost trades that are cheaper to elicit from privately informed sellers. Conversely, the trades removed by the algorithm have higher costs and higher values. Supporting these high-cost trades would require higher prices or larger information rents, so the algorithm curtails them and expands lower-cost trades even when their match values are lower.

\subsection{Market Segmentation}\label{sectionData}

Algorithmic recommendations also change the implication of price discrimination. 
Without an algorithm, more information about buyer values lets sellers tailor prices and typically changes both allocations and surplus. 
With an optimal algorithm, the effect is different: segmentation affects the distribution of prices across costs and values, but not the product allocation or the average division of surplus.

Market segmentation can be represented as an information structure $\info=(\signals,\map)$, where $\signals=\times_j\signals_j$ is a set of signal profiles, $\signal_j\in\signals_j$ is observed by seller $j$, and $\{\map(\cdot|\vecwtp)\}_{\vecwtp\in[0,1]^J}$ is a family of distributions over $\signals$.
The information structure is commonly known and signal realizations are independent of seller costs conditional on values.

We assume that the algorithm can distinguish market segments, so that sellers' costs remain the only information the algorithm does not observe.\footnote{Because the algorithm knows the value, this assumption is trivially satisfied if the sellers' signals are fully informative or, more generally, partitional.} Thus, the algorithm can condition its recommendations on realized signals, values, and prices. Market segmentation changes the information on which sellers condition their prices while preserving the cost-incentive problem.

For each information structure $\info$, we consider the $\weight$-optimal algorithm tailored to that structure.
Thus, as $\info$ varies, so does the corresponding optimal algorithm.

The signal-contingent version of the optimal algorithm is the direct analogue of \autoref{prop:competing-sellers}. Let $\tilde \signal_j$ denote seller $j$'s signal as a random variable. For each seller $j$, signal realization $\signal_j$, and active type $c_j$, define
\begin{equation}\label{eq:segmented-general-price}
p^*_{\weight,j}(c_j,\signal_j)
=
\E\!\left[
\virtual_{\weight,j}^{-1}(\theta_{\weight,j})
\mid
\theta_{\weight,j}\ge \virtual_{\weight,j}(c_j),\,
\tilde \signal_j=\signal_j
\right],
\end{equation}
where $\theta_{\weight,j}$ is defined in \eqref{eq:theta-alpha}.
At each profile of signals, values, and prices, the algorithm recommends a seller who maximizes the weighted virtual surplus among the types consistent with the observed prices and signals, as in \autoref{prop:competing-sellers}; the formal construction is in Appendix~\ref{app:segmented-construction}.

\begin{proposition}[Segmentation Neutrality]\label{propositionDataSeller-competing}
Across all market segmentations, buyer surplus, total surplus, each seller’s interim profit, and the ex post product allocation induced by the $\weight$-optimal algorithms are the same.
\end{proposition}
\noindent\emph{Proof.} See Appendix~\ref{app:segmented-construction}.\hspace*{\fill}$\square$\par\smallskip

The proof of \autoref{propositionDataSeller-competing} extends \autoref{lemmaImplementation} to the case in which sellers observe signals about the buyer's values, then establishes the equivalence between the $\weight$-optimal direct mechanism and the $\weight$-optimal algorithm under any information structure.
Once the algorithm accounts for the information available to sellers, market segmentation does not change the product allocation or division of surplus.

At the same time, finer market segmentation changes equilibrium behavior and the resulting price dispersion. We say that $\info_H$ is more informative than $\info_L$ for seller $j$ if seller $j$'s marginal signal under $\info_H$ is Blackwell more informative about $\vecwtp$ than that under $\info_L$ \citep{blackwell1951comparison}.
Under an information structure $\info$, the \emph{transaction-price distribution} of an active type $\type_j$ is the distribution of the equilibrium price $p^*_{\weight,j}(\type_j,\tilde\signal_j)$ in \eqref{eq:segmented-general-price} conditional on the buyer trading with seller $j$.\footnote{That is, conditional on $\theta_{\weight,j}\ge\virtual_{\weight,j}(\type_j)$; the proof of \autoref{propositionDataSeller-competing} shows that the probability of this event does not depend on $\info$.}

\begin{proposition}[Segmentation and Price Dispersion]\label{propositionPartitionalcompeting}

If $\info_H$ is more informative than $\info_L$ for seller $j$, then for every active type of seller $j$, the transaction-price distribution under $\info_H$ is a mean-preserving spread of that under $\info_L$. If the comparison holds for every seller, then the distribution of the price the buyer pays conditional on trade under $\info_H$ is a mean-preserving spread of that under $\info_L$.

\end{proposition}

\noindent\emph{Proof.} See Appendix~\ref{app:proof-price-dispersion}.\hspace*{\fill}$\square$\par\smallskip

\autoref{propositionPartitionalcompeting} implies that finer market segmentation does not change the average of transaction prices but makes them more dispersed.
Fixing an active type, the equilibrium price $p^*_{\weight,j}(c_j,\signal_j)$ in \eqref{eq:segmented-general-price} is the expectation conditional on the event that seller $j$ transacts with the buyer.
Conditional on this event, a more informative signal refines seller $j$'s posterior, and while the posterior mean remains the same, the realized conditional means---that is, the prices---become more variable.
\addtocounter{example}{-1}
\begin{example}[Uniform, continued]
	We continue with the uniform example and compare no segmentation with full segmentation.\footnote{Appendix~\ref{app:uniform-segmentation} derives the corresponding equilibrium prices.}
	\autoref{fig:segmentation} depicts the optimal algorithm thresholds (left) and, for $\weight=1$ and $\type=1/4$, transaction prices (right).
	For the seller type $\type=1/4$ shown in the right panel, trade occurs under either segmentation if and only if $\wtp\ge1/2$.
	With no segmentation, the seller charges $3/8$; with full segmentation, she charges $\wtp/2$.
	As a result, full segmentation lowers the price for $\wtp<3/4$ and raises it for $\wtp>3/4$, while preserving the average transaction price and welfare outcomes.
\begin{figure}
	\centering
	\begin{tikzpicture}
		\begin{axis}[
		width=0.45\textwidth,
		height=0.45\textwidth,
		    xmin=0, xmax=1.1,
		    ymin=0, ymax=1.1,
		    axis lines=middle,
		    xtick={0.5,1},
		    ytick={0.5,1},
		    xlabel={$p$},
		    ylabel={$v$},
		    xlabel style={at={(ticklabel* cs:1)}, anchor=north west, shift={(0ex,0ex)}},
		    ylabel style={at={(ticklabel* cs:1)}, anchor=south east, shift={(0ex,0ex)}},
		]
		    \addplot[very thick,black] coordinates {(0,0) (1/4,0) (1/2,1) (1,1)};
		    \addplot[very thick,dotted,black] coordinates {(0,0) (1/2,1) (1,1)};
		    \node[black] at (axis cs:0.23,0.84) {$\weight=1$};
		\end{axis}
	\end{tikzpicture}
	\hspace{0.5cm}
	\begin{tikzpicture}
	    \begin{axis}[
		width=0.47\textwidth,
		height=0.45\textwidth,
		xmin=0, xmax=1.1,
		ymin=0, ymax=0.6,
		axis lines=middle,
		xtick={0.5,0.75,1},
		xticklabels={$\tfrac{1}{2}$,$\tfrac{3}{4}$,$1$},
		ytick={0.25,0.375,0.5},
		yticklabels={$\tfrac{1}{4}$,$\tfrac{3}{8}$,$\tfrac{1}{2}$},
		xlabel={$\wtp$},
		ylabel={$p$},
		xlabel style={at={(ticklabel* cs:1)}, anchor=north west, shift={(0ex,0ex)}},
		ylabel style={at={(ticklabel* cs:1)}, anchor=south east, shift={(0ex,0ex)}},
	    ]
		\addplot[very thick,black] coordinates {(0.5,0.375) (1,0.375)};
		\addplot[very thick,dotted,black] coordinates {(0.5,0.25) (1,0.5)};
		\draw[dashed,black!50] (axis cs:0.5,0) -- (axis cs:0.5,0.55);
		\node[black!65] at (axis cs:0.25,0.08) {no trade};
	    \end{axis}
	\end{tikzpicture}
	\caption{Left: Algorithm thresholds under no segmentation (solid) and full segmentation (dotted) for $\weight=1$. Right: transaction prices for $\type=1/4$ under no segmentation (solid) and full segmentation (dotted). $J=1$, $\wtp\sim U[0,1]$, $\type\sim U[0,1]$.}
	    \label{fig:segmentation}
	\end{figure}
\hfill$\closing$
\end{example}

The economic lesson is that the welfare consequences of market segmentation depend on the buyer's technology. If buyers use algorithmic recommendations, sellers charge higher prices in favorable segments and lower prices in unfavorable segments, while the allocation and the relevant average surplus objects remain unchanged. This qualifies the usual concern about personalized pricing. 
The issue is whether buyer decisions are mediated by a recommendation technology that can respond to the sellers' information. 
The design of the algorithm is essential to this argument.
For example, if a monopoly seller has full information about the buyer's value, a myopic algorithm allows perfect price discrimination, whereas the optimal algorithm preserves average buyer surplus.

\section{Extensions}\label{sec:discussion}

\subsection{Gatekeeping Relaxation}\label{sec:roesler}

Up to this point, we have assumed that the buyer cannot purchase a product if it is not recommended.
An alternative setting is one in which the buyer knows of the product's existence but not its match value.
To study such a setting, we now allow the buyer to purchase a product even if it is not recommended.
For clarity and to facilitate comparison with the existing literature, we focus on the case of a single seller, $J=1$.
The objective of the algorithm remains $\Wobj_\weight$ in \eqref{eq:main-objective} with an arbitrary $\weight\in[0,1]$.

First, as in \autoref{sec:known_cost}, we depart from the baseline model and assume that the seller's cost is commonly known to be $c_0$.  This case is closest to \cite{roesler2017buyer} and, when $\alpha=1$, differs only in the timing of the recommendations. 
 In their setting, recommendations come before the price is posted and thus cannot condition on the price; in our setting, the recommendations can condition on the price.

By \autoref{claim:known_cost}, when the cost is known to be $c_0$,  an optimal algorithm with gatekeeping recommends the product if and only if $ p=\type_0 $ and $ v\geq \type_0 $: if the seller sets $ p=\type_0 $, the outcome is efficient and leaves the seller with zero profit. If $ \expect[\wtp] \le c_0 $, an algorithm can attain this outcome even without gatekeeping by revealing no information whenever $p > \type_0$.

In contrast, if $ \expect[\wtp]> c_0 $, the seller can secure a positive profit by charging a price $p \in (\type_0, \expect[\wtp])$, because no algorithm can make the buyer believe that the expected value of the product is always below $p< \expect[\wtp]$.
In this case, the optimal algorithm deters the seller from setting a higher price via \emph{adversarial persuasion}: at each price, the algorithm provides the buyer with information that minimizes the probability of trade.
Specifically, the algorithm reveals whether $ \wtp < \hat{\wtp}(p)$, where $\hat{\wtp}(p)$ is such that 
\begin{align} \expect[\tilde\wtp|\tilde\wtp < \hat{\wtp}(p)]=\min\{p,\expect[v]\},
\end{align}
and persuades the buyer to purchase only when $\wtp \ge \hat{\wtp}(p)$.
The maximum profit the seller can guarantee against adversarial persuasion is 
\begin{equation}\label{equationMinimal}
\underline\profit \triangleq \max_{p \ge \type_0} (p-\type_0)[1 - \wtpmeas(\hat{\wtp}(p))].
\end{equation}
The optimal algorithm induces an efficient trade while leaving the seller with profit $\underline\profit$.

\begin{proposition}[No Gatekeeping. Known Cost]\label{propositionDisclosure}
Suppose that the seller's cost is commonly known to be $\type_0\in [0,1)$,  $\wtpmeas$ is continuous with full support on $[0,1]$, and  the buyer can purchase the product even when not recommended.  For every $\weight\in[0,1]$, the $\weight$-optimal algorithm recommends the product if $p=p^*\triangleq\type_0+  \underline{\profit}/(1 - \wtpmeas(\type_0))$ and $v\geq \type_0$ and uses adversarial persuasion at all other prices. Under this algorithm, the seller posts a price $p^*$, and the equilibrium trade is efficient.
\end{proposition}

\noindent\emph{Proof.} See Appendix \ref{proofpropositionDisclosure}.\hspace*{\fill}$\square$\par\smallskip

As in \cite{roesler2017buyer}, the buyer-optimal algorithm leads to efficient trade. However, the seller's rent is driven by adversarial persuasion price-by-price and thus is lower, reaching zero when $\expect[\wtp]<\type_0$.
Therefore, if and only if $\type_0 < \expect[\wtp]$, the design and consequences of the optimal algorithm depend on whether the buyer can purchase the nonrecommended product.

We now turn to the case in which the seller's costs are uncertain.
In this case, the optimal algorithms with and without gatekeeping can coincide, even when the cost can fall below $\expect[\wtp]$.
This happens, for example, in the uniform setting of \autoref{exampleUniform}: under the buyer-optimal algorithm, the 
buyer has no incentive to purchase the product after receiving no recommendation, even if she could do so, because the expected value of the product conditional on no recommendation is always below the price.
The following result provides a sufficient condition under which this occurs.

\begin{proposition}[No Gatekeeping. Unknown Cost]\label{claimDisclosure}
Suppose that $\wtpmeas$ is continuous with full support on $[0,1]$. Let $\weight \in [0,1]$. If $\int^{\virtual_\weight(\type)}_0 [\wtp - \type] \dint\wtpmeas(\wtp) \leq 0$ for each $\type \in [0,\overline c_\weight]$, then even if the buyer can purchase the product when not recommended, the $\weight$-\algo in \autoref{prop:single-weight-optimal} is $\weight$-optimal.
\end{proposition}

\noindent\emph{Proof.} See Appendix \ref{proofclaimDisclosure}.\hspace*{\fill}$\square$\par\smallskip

The condition of \autoref{claimDisclosure} ensures that whenever the product is not recommended under the algorithm of \autoref{prop:single-weight-optimal}, the buyer infers that the expected value of the product is below the price. The condition at $\weight=1$ implies the condition for every $\weight\in[0,1]$. It holds, for example, when (i) $\typemeas(\type) = \type^x$ and $\wtpmeas(\wtp) = \wtp^y$ with $0<y\leq x$  or (ii) $\wtpmeas$ is uniform and $\typemeas(\type)/\type$ is increasing.\footnote{Point (i) follows from direct calculation. For Point (ii), note that the condition $( \typemeas(\type)/\type)' \ge 0$ can be written as $\virtual(\type)/\type\le 2$, which is equivalent to $\int^{\virtual(\type)}_0 [\wtp - \type] \dint\wtpmeas(\wtp) \le  0$ when $G  = U[0,1]$.
}

To see the intuition behind this condition, suppose that the seller posts a price of $\popt_\weight(\type)$, and the \algo recommends that the buyer not purchase the product, which, by \autoref{prop:single-weight-optimal}, reveals that $\wtp \le \virtual_\weight(\type)$.
If the buyer purchases the product, the seller's profit increases, but total surplus decreases because $\int^{\virtual_\weight(\type)}_0 [\wtp - \type] \dint\wtpmeas(\wtp) \le  0$.
Then, the buyer's payoff must decrease, and thus she follows the recommendation not to buy.
Thus, in some cases, gatekeeping can be fully replaced by persuasion, whereas in others it provides additional control.

\subsection{Seller-Aligned  Objectives}\label{sectionSellerSurplusObjective}

The main analysis focuses on objectives between buyer surplus and total surplus. Additional tools are required once the algorithm instead maximizes a weighted average of total surplus and seller surplus,
\begin{equation}\label{eq:seller-objective}
(1-\sellerweight)\TS+\sellerweight\SSurplus,
\end{equation}
where $\sellerweight\in[0,1]$. This objective is relevant for seller-sponsored platforms or intermediaries whose incentives are better aligned with sellers. 

In Appendix~\ref{app:seller-surplus-deterministic-values}, we characterize the optimal algorithm for any $\sellerweight$ when the value profile is commonly known.
Here, we focus on the seller-optimal algorithm, $\sellerweight=1$, and the common-value benchmark, which highlights the contrast with the baseline optimal algorithm most clearly.

\begin{proposition}[Seller-Optimal Algorithm]\label{propAlphaOptimalSymmetric}
Suppose all sellers have the same deterministic value $v\in(0,1]$, costs are i.i.d. with common distribution $F$ and density $f$, and $F/f$ is strictly increasing on $[0,1]$. 
Then, a seller-optimal algorithm recommends uniformly among sellers who post $v$ and does not recommend sellers who post any other price.
\end{proposition}

\noindent\emph{Proof.} See Appendix \ref{app:seller-surplus-deterministic-values}.\hspace*{\fill}$\square$\par\smallskip

The buyer-aligned algorithm favors lower-cost sellers and punishes high prices to lower equilibrium prices.
In contrast, the seller-optimal algorithm encourages high prices and pools all trades at the buyer's value.
In equilibrium, all sellers with $c_j\le v$ post price $p_j=v$, whereas sellers with $c_j>v$   never trade. Thus, the equilibrium is cost-inefficient.

When $J\geq 2$ and the buyer's values are uncertain, the contrast with the main analysis is even sharper. 
\autoref{claimMechanismGapSellerObjective} in Appendix~\ref{app:seller-surplus-mechanism-gap} shows that the objective value under the optimal algorithm is typically lower than that under the optimal mechanism that imposes the buyer's ex ante individual rationality constraint.
Intuitively, a direct mechanism can maximize seller profits by implementing the efficient allocation, charging the buyer a lump-sum fee, and distributing it to the sellers.
In this way, the mechanism maximizes total surplus without leaving any surplus to the buyer.
A recommendation algorithm cannot administer such transfers, so it either has to induce an inefficient allocation or leave the buyer a positive surplus, either of which leads to a lower seller surplus.\footnote{A similar issue arises in \cite{muto2020revenue}, who maximize auction surplus subject to a cap on the auctioneer's expected revenue. There, a lump-sum payment from the auctioneer to bidders would avoid distorting the allocation, and they study the consequences of excluding this instrument.}
Characterizing seller-aligned algorithms would require different tools, and we leave a general characterization for future research.

\section{Conclusion}
In this paper, we studied the design of algorithmic recommendations in the presence of strategic pricing. The main lesson is that algorithmic recommendations can substitute for more intrusive instruments of mechanism design. They can discipline prices, partly substitute for ordinary seller competition, reshape the composition of trade, and absorb the average effects of market segmentation.

We view our work as a stepping stone toward a better understanding of algorithmic design in strategic settings, an area of growing importance at the intersection of economics and computer science (e.g., \citealt{goktas2025strategic}). First, our model of algorithms is deliberately stylized to analyze strategic motives in a clear and tractable way. A practical implementation would ideally incorporate engineering concerns from which we abstract, such as the details of value estimation, computational complexity, and robustness. Second, it would be interesting to study market structures for algorithm providers and understand which of the algorithms that we characterize are favored by different market structures. Third, the ideas developed here about algorithmic decision-making can be extended beyond consumption settings---for example, to algorithmic matching or algorithmic negotiations. All of this future research can build on the analytical framework developed in this paper.

\bibliography{algo_revision}
\bibliographystyle{ecta}

\appendix
\section{Proofs}\label{appendix}

\subsection{Proof of \autoref{claim:known_cost}}\label{app:proof-known-cost}
\begin{proof}
For each value profile $\vecwtp$, let
\begin{equation}\label{algoeff}
M(\vecwtp)\triangleq \argmax_{k\in\Jzero}\{\wtp_k-c_k\}.
\end{equation}
Consider an algorithm that recommends no purchase if $0\in M(\vecwtp)$. If $0\notin M(\vecwtp)$, it recommends a seller in $M(\vecwtp)$ whose price equals her cost, with ties broken arbitrarily, and recommends no purchase if no such seller exists.

Consider the strategy profile in which every seller sets $p_j=c_j$ and the buyer follows every recommendation. Every recommended seller $j$ satisfies $\wtp_j-p_j>0$, so the buyer is willing to follow. Each seller earns zero profit. A deviation to any $p_j\ne c_j$ makes seller $j$ ineligible for recommendation and also yields zero profit, so no seller has a profitable deviation. Thus, these strategies form an equilibrium. By \eqref{algoeff}, the resulting allocation is efficient and every seller earns zero profit.

Let $S^*\triangleq\E[\max_{k\in\Jzero}\{\wtp_k-c_k\}]$ denote the expected total surplus under the efficient allocation. The constructed equilibrium attains $\TS=\BS=S^*$. Under any algorithm and equilibrium, $\TS\le S^*$, and sellers' equilibrium profits are nonnegative because each seller can guarantee zero profit by setting her price equal to her cost. Hence,
\[
\Wobj_\weight=\TS-\weight\SSurplus\le S^*
\]
for every $\weight\in[0,1]$. The constructed outcome attains this bound.
\end{proof}
\subsection{Proof of \autoref{lemmaImplementation}}\label{app:proof-implementation-lemma}
\begin{proof}
Let $(q_j,t_j)_{j\in\Jset}$ be the ex post allocation and transfer rules of the direct mechanism. For each seller $j$, define $U_j(c_j)=T_j(c_j)-c_jQ_j(c_j)$.
Conditions 1 and 2 imply $U_j(c_j)=\int_{c_j}^1 Q_j(x)\,\dint x$ and the standard incentive compatibility,
\begin{equation*}
U_j(c_j)\ge T_j(c'_j)-c_jQ_j(c'_j)
\quad\text{for all }c_j,c'_j.
\end{equation*}
Define the candidate price map
\begin{equation*}
    p_j^*(c_j) = \begin{cases} T_j(c_j) / Q_j(c_j) & \text{for } Q_j(c_j) > 0, \\ 2 & \text{for } Q_j(c_j) = 0. \end{cases}
\end{equation*}
For any type with $Q_j(c_j)>0$, Condition 3 and values in $[0,1]$ imply $T_j(c_j)\le V_j(c_j)\le Q_j(c_j)$, and therefore $p_j^*(c_j)\le1$. Thus the price $2$ is inactive.

We first record two properties of the price map. The function $p_j^*(c_j)$ is weakly increasing.
Let $c'$ and $c''$ be types that satisfy $c' < c''$ and $Q_j(c'') > 0$. The difference in prices equals
\begin{equation}\label{equationPP}
    p_j^*(c'') - p_j^*(c') = \frac{(c'' - c')\, Q_j(c') - \int_{c'}^{c''} Q_j(x)\, dx}{Q_j(c')} + \left(\frac{1}{Q_j(c'')} - \frac{1}{Q_j(c')}\right) \int_{c''}^{1} Q_j(x)\, dx.
\end{equation}
Condition 1 implies $Q_j$ is nonincreasing in $c_j$. 
The inequality $\int_{c'}^{c''} Q_j(x)\, dx \leq (c'' - c')\, Q_j(c')$ ensures the first term is nonnegative. The inequality $Q_j(c'') \leq Q_j(c')$ ensures the second term is nonnegative. 
Thus, the sum is nonnegative.
When $Q_j(c'')=0$, we have $p_j^*(c'')=2$, while $p_j^*(c')\le2$ by the previous paragraph. Thus $p_j^*$ is weakly increasing on $[0,1]$.

Second, $p_j^*(c') = p_j^*(c'')$ implies that $Q_j(c)$ is constant on the interval $(c', c'')$.
Suppose $p_j^*(c') = p_j^*(c'')$ for types $c' < c''$. 
If $p_j^*(c'') = 2$, we must have $Q_j(c') = Q_j(c'') = 0$. 
This implies $Q_j(c) =0$ for any $c \in (c', c'')$.
If $p_j^*(c'') < 2$, we have $Q_j(c'') > 0$.
Because $p_j^*(c'') = p_j^*(c')$, both nonnegative terms on the RHS of \eqref{equationPP} are $0$. 
The first term equaling zero implies that $Q_j$ is constant on the interval $(c', c'')$. 

Define $C_j(p) = \{c_j \in [0,1] \mid p_j^*(c_j) = p\}$. 
Because $p^*_j(c_j)$ is nondecreasing, $C_j(p)$ is an interval.
The argument in the previous paragraph then implies that $Q_j$ is constant on the interior of $C_j(p)$. 
Let $\bar{Q}_j(p)$ denote this constant when $C_j(p)$ has nonempty interior; when $C_j(p)=\{c_j\}$ is a singleton, set $\bar{Q}_j(p)=Q_j(c_j)$.
Let $N_j$ be the set of boundary points of nondegenerate intervals $C_j(p)$. 
This set is countable and thus has $\typemeas_j$-measure zero.

We now construct the algorithm.
Define the recommendation algorithm as
\begin{equation*}
    \alg_j(\mathbf{v}, \mathbf{p}) = \mathbb{E}_{\mathbf{c}}\!\left[q_j(\mathbf{v}, \mathbf{c}) \;\middle|\; c_k \in C_k(p_k) \text{ for all } k\right]
\end{equation*}
for price profiles $\mathbf{p}$ that satisfy $p_k \in \mathrm{Image}(p_k^*(\cdot))$ for all $k$. For price profiles $\mathbf{p}$ with $p_k \notin \mathrm{Image}(p_k^*)$ for some $k$, set $\alg_j(\mathbf{v}, \mathbf{p}) = 0$ for all $j$.  The conditional expectation is taken with respect to the product of the posterior type distributions induced by the price profile; if $C_k(p_k)$ is a singleton, the posterior for seller $k$ is degenerate at that type.
Linearity of expectation yields $\sum_{j=1}^J \alg_j(\mathbf{v}, \mathbf{p}) \leq 1$.
Set $\alg_0(\mathbf{v},\mathbf{p})=1-\sum_{j=1}^J\alg_j(\mathbf{v},\mathbf{p})$.

\emph{Buyer obedience:} Consider a price $p_j \in \mathrm{Image}(p_j^*)$ at which seller $j$ is recommended with positive probability, so $\bar Q_j(p_j)>0$. Upon receiving recommendation $r = j$ at price $p_j$, the buyer infers $c_j \in C_j(p_j)$. 
Condition 3 ensures $V_j(c_j) \geq T_j(c_j)$. 
The definition of $p_j^*$ implies $T_j(c_j) = p_j\, Q_j(c_j)$ for $Q_j(c_j) > 0$. This equals $p_j \bar{Q}_j(p_j)$ at the unique type when $C_j(p_j)$ is a singleton and for almost every type in $C_j(p_j)$ when $C_j(p_j)$ is nondegenerate. 
All expectations over $c_j\in C_j(p_j)$ in this paragraph use the posterior induced by the price before conditioning on the recommendation.
The probability of recommendation $r=j$ conditional on price $p_j$ equals $\mathbb{E}_{c_j \in C_j(p_j)}[Q_j(c_j)]=\bar{Q}_j(p_j)$, and the corresponding expected gross value equals $\mathbb{E}_{c_j \in C_j(p_j)}[V_j(c_j)]$.
Therefore $\mathbb{E}_{c_j \in C_j(p_j)}[V_j(c_j)] \geq p_j\, \bar{Q}_j(p_j)$. Division by the recommendation probability $\bar{Q}_j(p_j) > 0$ yields $\mathbb{E}[v_j \mid r = j,\, p_j] \geq p_j$. 
The expected payoff of the buyer from a purchase is nonnegative.

\emph{Seller equilibrium:} Seller $j$'s strategy is to post $p_j^*(c_j)$ for every type $c_j$.
We first verify that this price attains payoff $U_j(c_j)$ for every type, including boundary points of pooling intervals. Fix $c_j$ and set $p=p_j^*(c_j)$. If $C_j(p)=\{c_j\}$ and $Q_j(c_j)>0$, then
$$
(p-c_j)\bar Q_j(p)
=T_j(c_j)-c_jQ_j(c_j)
=U_j(c_j).
$$
If $C_j(p)=\{c_j\}$ and $Q_j(c_j)=0$, monotonicity of $Q_j$ implies $Q_j(x)=0$ for all $x\ge c_j$, so $U_j(c_j)=0=(p-c_j)\bar Q_j(p)$. Finally, suppose $C_j(p)$ is nondegenerate. If $\bar Q_j(p)=0$, then $p=2$ and $Q_j(c)=0$ for every $c\in C_j(p)$, so the same equality holds. If $\bar Q_j(p)>0$, then every interior point $x$ of $C_j(p)$ satisfies $Q_j(x)=\bar Q_j(p)$ and $T_j(x)=p\bar Q_j(p)$, so
$$
(p-x)\bar Q_j(p)=U_j(x).
$$
Because $Q_j$ is bounded, $U_j(c)=\int_c^1 Q_j(x)\,\dint x$ is continuous in $c$. Therefore the preceding equality extends from the interior of $C_j(p)$ to its boundary points. Combining the singleton and nondegenerate cases gives
\begin{equation}\label{eq:boundary-payoff-identity}
(p_j^*(c_j)-c_j)\bar Q_j(p_j^*(c_j))=U_j(c_j)
\quad\text{for every }c_j.
\end{equation}

We show no seller has a profitable deviation. For any on-path price $p$, choose a representative type $\hat c_j(p)\in C_j(p)$ such that $\bar Q_j(p)=Q_j(\hat c_j(p))$ and $p\,\bar Q_j(p)=T_j(\hat c_j(p))$; for nondegenerate $C_j(p)$, choose any interior type, and for singleton $C_j(p)$, choose its only element. A type $c_j$ that deviates to price $p$ obtains
\begin{equation*}
p\,\bar Q_j(p)-c_j\bar Q_j(p)
=T_j(\hat c_j(p))-c_jQ_j(\hat c_j(p))
\le U_j(c_j),
\end{equation*}
where the inequality is the incentive inequality implied by Conditions 1 and 2. Deviations to prices outside the image of $p_j^*$ yield zero payoff, which is no more than $U_j(c_j)$. By \eqref{eq:boundary-payoff-identity}, posting $p_j^*(c_j)$ gives type $c_j$ payoff $U_j(c_j)$, so the proposed strategies form a profile of seller best responses. Together with buyer obedience, they form a perfect Bayesian equilibrium.

\emph{Seller interim profits, buyer surplus, and total surplus:}
Equation \eqref{eq:boundary-payoff-identity} establishes that every seller type obtains the same interim profit as under the direct mechanism. For every $c_j\notin N_j$, the construction and the constancy of $Q_j$ on the interior of each price preimage imply that seller $j$'s induced interim allocation equals $Q_j(c_j)$. At types in $N_j$, the induced interim allocation may differ from $Q_j(c_j)$, but these types form a null set.

For every value profile $\vecwtp$ and seller $j$, the law of iterated expectations gives
\begin{equation*}
\mathbb{E}_{\vectype}\!\left[
\alg_j(\vecwtp,p_1^*(c_1),\ldots,p_J^*(c_J))
\right]
=
\mathbb{E}_{\vectype}\!\left[q_j(\vecwtp,\vectype)\right].
\end{equation*}
Thus the expected gross value generated by trade is the same as under the direct mechanism. Expected production cost is also preserved because the interim allocations $(Q_j)_{j\in\Jset}$ are preserved almost everywhere. Hence total surplus is preserved. Integrating \eqref{eq:boundary-payoff-identity} over types and summing across sellers shows that seller surplus is preserved. Therefore, the identity $\TS=\BS+\SSurplus$ implies that buyer surplus is preserved as well.

Conversely, suppose $(Q_j,T_j,V_j)_{j\in\Jset}$ is induced by an algorithm and an equilibrium. For each seller $j$, let $U_j(c_j)=T_j(c_j)-c_jQ_j(c_j)$
be seller $j$'s equilibrium payoff. 
Because type $c_j$ can imitate the equilibrium strategy of type $c'_j$, we have  
\begin{equation*}
U_j(c_j)\geq T_j(c'_j)-c_jQ_j(c'_j)
\quad\text{for all }c_j,c'_j.
\end{equation*}
The standard envelope argument then implies that $Q_j$ is nonincreasing and
\begin{equation*}
U_j(c_j)=U_j(1)+\int_{c_j}^{1}Q_j(x)\,\dint x.
\end{equation*}
Because values are bounded above by $1$, a seller of cost $1$ cannot earn positive profit at any price at which the buyer is willing to purchase.
Hence $U_j(1)=0$, and Conditions 1 and 2 follow.

It remains to verify Condition 3. 
Take any equilibrium and fix seller $j$. 
For any price $p$, let $\bar q_j(p)$ denote the probability that the buyer purchases product $j$ conditional on that  
seller $j$ posts $p$, where the probability is over values, the other sellers'
prices, and recommendation events.
Let $\bar v_j(p)$ denote the expectation of $\wtp_j$ multiplied by
the same purchase indicator. 
Let $\sigma_{c_j}$ denote the equilibrium price distribution of type $c_j$.
We have
\begin{equation*}
T_j(c_j)=\int p\,\bar q_j(p)\,\dint\sigma_{c_j}(p)
\qquad\text{and}\qquad
V_j(c_j)=\int \bar v_j(p)\,\dint\sigma_{c_j}(p).
\end{equation*}
The buyer obedience implies that $\bar v_j(p)\ge p\,\bar q_j(p)$ if $p$ leads to purchase, and $\bar v_j(p)=\bar q_j(p)=0$ otherwise.
Integrating this inequality using $\sigma_{c_j}$ gives $V_j(c_j)\ge T_j(c_j)$.
\end{proof}

\subsection{Proof of \autoref{prop:competing-sellers}}\label{app:proof-competing-sellers}\label{app:proof-algorithm-mechanism-equivalence}
\begin{proof}
Let $U_j(c_j)=T_j(c_j)-c_jQ_j(c_j)$ denote seller $j$'s interim profit.
For $\weight>0$, optimality requires $U_j(1)=0$, while for $\weight=0$, $U_j(1)=0$ follows from the normalization in \autoref{def:alpha-optimal}.
Hence, the envelope formula gives $U_j(c_j)=\int_{c_j}^1Q_j(x)\,\dint x$, which implies 
\begin{equation*}
\mathbb E[U_j(c_j)]
=\mathbb E\!\left[Q_j(c_j)\frac{\typemeas_j(c_j)}{\typedens_j(c_j)}\right].
\end{equation*}
Therefore, for any normalized direct mechanism,
\begin{equation}\label{eq:objective-virtual-surplus}
\Wobj_\weight
=
\mathbb E\!\left[
\sum_{j\in\Jset}q_j(\vecwtp,\vectype)
\left(\wtp_j-\virtual_{\weight,j}(c_j)\right)
\right].
\end{equation}
The same representation applies to any algorithm and equilibrium.\footnote{Namely, we obtain the same expression after we define $q_j(\vecwtp,\vectype)$ as the equilibrium probability that seller $j$ is recommended and purchased at $(\vecwtp,\vectype)$, where the definition integrates over any equilibrium price mixing.}

Equation \eqref{eq:objective-virtual-surplus} is bounded above pointwise by
\begin{equation*}
\mathbb E\!\left[
\max\left\{0,\max_{j\in\Jset}\left(\wtp_j-\virtual_{\weight,j}(c_j)\right)\right\}
\right].
\end{equation*}
The direct mechanism that allocates the product to a seller who attains this maximum whenever the maximum is positive attains the bound; the allocation is monotone in each seller's own cost by \autoref{ass:regular-virtual-costs}, and the transfer schedule is pinned down by the envelope formula and the normalization $U_j(1)=0$. Hence every $\weight$-optimal direct mechanism allocates only to generalized-virtual-surplus maximizers, except possibly on tie sets.
Tie-breaking on these sets is immaterial for ex ante payoffs and for the almost-everywhere allocation conclusion below.
Fix one such direct mechanism and let $Q_j^D$ and $T_j^D$ denote its interim allocation and revenue schedules.

We next verify that this optimal direct mechanism satisfies the conditions of \autoref{lemmaImplementation}. Conditions 1 and 2 follow from the envelope formula. To verify Condition 3, fix seller $j$, type $c_j$, and a realization $(\vecwtp,\vectype_{-j})$. Define
\begin{equation*}
r_j(x)=q_j(\vecwtp,x,\vectype_{-j}).
\end{equation*}
The optimal allocation rule implies $r_j(x)=0$ whenever $x>\wtp_j$, because seller $j$ can be allocated only if $\wtp_j\ge \virtual_{\weight,j}(x)\ge x$. It also implies that $r_j(x)$ is weakly decreasing in $x$. Thus, pointwise in $(\vecwtp,\vectype_{-j})$,
\begin{equation*}
\int_{c_j}^1 r_j(x)\,\dint x
\le
(\wtp_j-c_j)r_j(c_j),
\end{equation*}
where the inequality is trivial when $r_j(c_j)=0$, and otherwise follows because $r_j(x)=0$ for $x>\wtp_j$ and $r_j(x)\le r_j(c_j)$ for $x\ge c_j$. Taking expectations gives
\begin{equation*}
c_jQ_j(c_j)+\int_{c_j}^1Q_j(x)\,\dint x\le V_j(c_j).
\end{equation*}
Because $T_j(c_j)=c_jQ_j(c_j)+\int_{c_j}^1Q_j(x)\,\dint x$, Condition 3 follows. \autoref{lemmaImplementation} then gives an algorithm and equilibrium that attain the same buyer surplus, total surplus, and seller interim profits as the $\weight$-optimal direct mechanism. Hence an $\weight$-optimal algorithm attains the same value of $\Wobj_\weight$.

It remains to compare ex post allocations. Under \autoref{ass:regular-virtual-costs}, ties in generalized virtual surplus have probability zero and the generalized-virtual-surplus maximizer is unique almost surely. 
Any $\weight$-optimal direct mechanism and any $\weight$-optimal algorithm must attain the pointwise bound in \eqref{eq:objective-virtual-surplus} almost surely.
Hence they induce the same ex post product allocation for almost every $(\vecwtp,\vectype)$.

Because the ex post allocations coincide almost surely, total surplus is the same. Seller interim profits are also the same almost everywhere by the envelope formula and the common normalization at type $1$. 
Thus, buyer surplus, which is total surplus net of seller profits, is the same as well.

It remains to compute the prices in the implementation. Write $\virtual_j$ for $\virtual_{\weight,j}$, and define
\begin{equation*}
M_j=\max_{k\in\Jzero\setminus\{j\}}\{\wtp_k-\virtual_k(\type_k)\},
\end{equation*}
so that $\theta_{\weight,j}=\wtp_j-M_j$ by \eqref{eq:theta-alpha}. The verification above does not use the tie-breaking rule. Hence \autoref{lemmaImplementation} gives an algorithm and an equilibrium that implement the same interim outcomes as the fixed $\weight$-optimal direct mechanism. In that construction, every active type posts
\begin{equation*}
\hat p_j(c_j)=\frac{T_j^D(c_j)}{Q_j^D(c_j)},
\end{equation*}
including boundary types of pooling intervals, and every inactive type posts an inactive price. It remains to compute this ratio and to prove the recommendation claim.

Fix seller $j$. Define a random variable $Y_j$ by $Y_j=\virtual_j^{-1}(\theta_{\weight,j})$ when $\theta_{\weight,j}\ge0$, and set $Y_j=-1$ otherwise. The convention below zero is only notational: on the allocation event, $\theta_{\weight,j}\ge\virtual_j(c_j)\ge0$, so the inverse is evaluated on its natural domain. Because the dummy seller is included in $M_j$, $M_j\ge0$. Thus $\theta_{\weight,j}\le\wtp_j\le1$, and therefore $Y_j\le\virtual_j^{-1}(1)\le1$ whenever $Y_j\ge0$.

For every $c_j$, the direct allocation event for seller $j$ is, up to ties among real sellers,
\begin{equation*}
\wtp_j-\virtual_j(c_j)\ge M_j
\quad\Longleftrightarrow\quad
\theta_{\weight,j}\ge\virtual_j(c_j)
\quad\Longleftrightarrow\quad
Y_j\ge c_j.
\end{equation*}
Here, we break ties in favor of a real seller over the dummy seller; ties among real sellers have probability zero at every $c_j$, because the other sellers' costs have densities.
Consequently, $Q_j^D(c_j)=\Pr[Y_j\ge c_j]$ for every $c_j$.

Condition 2 of \autoref{lemmaImplementation} gives
\begin{equation*}
T_j^D(c_j)=c_jQ_j^D(c_j)+\int_{c_j}^{1}Q_j^D(x)\,\dint x.
\end{equation*}
For every realization of $Y_j$ and every $c_j$, we have
\begin{align*}
Y_j\,\ind\{Y_j\ge c_j\}
&=
c_j\,\ind\{Y_j\ge c_j\}
+\int_{c_j}^{\infty}\ind\{Y_j\ge x\}\,\dint x\\
\Rightarrow \E\!\left[Y_j\,\ind\{Y_j\ge c_j\}\right]
&=
c_jQ_j^D(c_j)+\int_{c_j}^{1}Q_j^D(x)\,\dint x
=T_j^D(c_j).
\end{align*}
For every active type $c_j$, division by $Q_j^D(c_j)>0$ yields
\begin{equation*}
\hat p_j(c_j)
=
\E[Y_j\mid Y_j\ge c_j]
=
\E\!\left[\virtual_j^{-1}(\theta_{\weight,j})\mid\theta_{\weight,j}\ge\virtual_j(c_j)\right]
=p^*_{\weight,j}(c_j),
\end{equation*}
which proves the first claim.

By the construction in \autoref{lemmaImplementation}, it remains only to show that, for almost every on-path profile $(\vecwtp,\vecp)$, the direct allocation is constant over all type profiles $\vectype$ that satisfy $c_j\in C_j(p_j)$ for every $j\in\Jset$. For any nondegenerate active price preimage and $c<c'$ in that preimage, equality in \autoref{equationPP}, together with $Q_j^D(x)=\Pr[Y_j\ge x]$, implies $\Pr[c\le Y_j<c']=0$; hence the threshold indicator is almost surely constant on the preimage. Singleton and inactive price preimages are immediate, and a finite union over sellers gives the claim. Therefore, the conditional expectation used by the algorithm is degenerate on the common maximizer $j^*(\vecwtp,\vecp)$, and the algorithm recommends this seller.
\end{proof}

\subsection{Proof of \autoref{prop:single-weight-optimal}}\label{app:proof-single-weight-optimal}
\begin{proof}
Write $\virtual$ for $\virtual_\weight$ and let $Y=\virtual^{-1}(\V)$, where $\V\sim \wtpmeas$ denotes the valuation as a random variable.
By the virtual-surplus representation in the proof of \autoref{propAlgorithmMechanismEquivalence}, an $\weight$-optimal direct mechanism allocates if and only if $v\ge \virtual(c)$.
For this allocation, we obtain 
\begin{equation*}
Q(c)=\Pr[Y\ge c] 
\end{equation*}
and
\begin{equation*}
T(c)=cQ(c)+\int_c^1 Q(x)\,\dint x.
\end{equation*}
Let $U(c)=T(c)-cQ(c)$ denote the seller's interim profit.
For any realization of $Y$ and a constant $c$, we have
$$
Y\,\ind\{Y\ge c\}
=
c\,\ind\{Y\ge c\}
+
\int_c^\infty \ind\{Y\ge y\}\,\dint y.
$$
Taking expectations yields
\begin{equation*}
\E\!\left[Y\,\ind\{Y\ge c\}\right]
=
c\Pr[Y\ge c]+\int_c^1 \Pr[Y\ge y]\,\dint y
=T(c).
\end{equation*}
The buyer-obedience condition in \autoref{lemmaImplementation} holds because
\begin{equation*}
\E\!\left[\V\,\mathds{1}\{Y\ge c\}\right]
\ge
\E\!\left[Y\,\mathds{1}\{Y\ge c\}\right]
=T(c),
\end{equation*}
where the inequality follows from $\V=\virtual(Y)\ge Y$.
Thus \autoref{lemmaImplementation} implements this direct mechanism.

For any active type with $Q(c)>0$, the price map in the proof of \autoref{lemmaImplementation} is
\begin{equation*}
p^*_\weight(c)=\frac{T(c)}{Q(c)}
=\E[Y\mid Y\ge c]
=\E\!\left[\virtual^{-1}(\V)\mid \V\ge \virtual(c)\right]
=\vst_\weight(\virtual(c)).
\end{equation*}
If $c>\overline c_\weight$, then $\virtual(c)>1$, so $Q(c)=0$ and the type is inactive.
The boundary type $c=\overline c_\weight$ is payoff-irrelevant when $Q(c)=0$; assign it the limiting active price or an inactive price.

It remains only to verify that the simple rule in the statement is a valid off-path extension of the implementation in \autoref{lemmaImplementation}.
The function $\vst_\weight$ is weakly increasing.
For each price $p$, the set
$$
R(p)=\{v:\vst_\weight(v)\ge p\}
$$
is an upper tail of the value space.
Let $q_p=\Pr[\V\in R(p)]$.
If $q_p=0$, a deviation to $p$ yields zero profit; specify off-path beliefs and a sequentially rational buyer action at any resulting null history.
Otherwise let $m_p=\E[Y\mid \V\in R(p)]$.
Because $R(p)$ is an upper tail, there exists a cutoff $z$ such that the event $\{\V\in R(p)\}$ agrees, up to the cutoff convention at a boundary atom, with either $\{Y\ge z\}$ or $\{Y>z\}$; equivalently, the distinction is whether the boundary value $\virtual(z)$ belongs to $R(p)$.
If $\virtual(z)\in R(p)$, then $\{\V\in R(p)\}=\{Y\ge z\}$ and $m_p=\E[Y\mid Y\ge z]\ge p$ by the definition of $R(p)$.
If $\virtual(z)\notin R(p)$, then $\{\V\in R(p)\}=\{Y>z\}$ and $m_p$ is the right limit of $\E[Y\mid Y\ge z']$ along cutoffs $z'>z$ for which $\{Y\ge z'\}\subset \{\V\in R(p)\}$; again $m_p\ge p$.
Moreover, because $\V=\virtual(Y)\ge Y$,
\begin{equation*}
\E[\V\mid \V\in R(p)]\ge m_p\ge p,
\end{equation*}
so the buyer is willing to purchase after every recommendation that occurs with positive probability at price $p$.
Thus a seller of type $c$ who deviates to price $p$ earns at most
\begin{equation*}
(m_p-c)q_p
=\E\!\left[Y\,\mathds{1}\{\V\in R(p)\}\right]-cq_p.
\end{equation*}
The same cutoff representation gives a sequence of reports $(c_n)$ with the desired limits: If $\virtual(z)\in R(p)$, take $c_n=z$ for all $n$; if $\virtual(z)\notin R(p)$, take any sequence $c_n\downarrow z$ with $c_n>z$. Then
$$
Q(c_n)\to q_p
\quad\text{and}\quad
T(c_n)\to \E\!\left[Y\,\mathds{1}\{\V\in R(p)\}\right].
$$
The direct mechanism's incentive constraints imply $T(c_n)-cQ(c_n)\le U(c)$ for every $n$.
Taking limits gives $(m_p-c)q_p\le U(c)$.
Therefore no off-path price is profitable.

Under the equilibrium price $p^*_\weight(c)=\vst_\weight(\virtual(c))$, the recommendation rule $\vst_\weight(v)\ge p^*_\weight(c)$ coincides with the direct allocation $\mathds{1}\{v\ge\virtual(c)\}$ on the support of $G$, with the cutoff included as in the direct allocation.
Indeed, if $\vst_\weight$ is flat below $\virtual(c)$, the corresponding interval has zero $G$-mass; otherwise adding those lower values to the tail would strictly lower the conditional mean.
Thus trade occurs if and only if $v\ge\virtual_\weight(c)$.
Because the implemented direct mechanism is $\weight$-optimal by \autoref{propAlgorithmMechanismEquivalence}, the algorithm and equilibrium described in the statement are $\weight$-optimal.
\end{proof}

\subsection{Proof of \autoref{prop:payoff-set}}\label{app:proof-payoff-set}

\begin{proof}
As we argued in the main text, when $\typemeas/\typedens$ is nondecreasing, $\virtual_\weight$ is strictly increasing for every $\weight\ge0$, and \autoref{prop:single-weight-optimal} extends verbatim to any $\weight\ge0$, including $\weight>1$ (except for the $\weight$-optimality claim). The resulting $\weight$-algorithm induces trade if and only if $\wtp\ge\virtual_\weight(\type)$. Define $K\triangleq\{(0,0)\}\cup\{(\BS_\weight,\SSurplus_\weight)\mid\weight\ge0\}$. Then $K\subseteq\payoffset$: The $\weight$-algorithms attain $(\BS_\weight,\SSurplus_\weight)$, and the algorithm that never recommends attains $(0,0)$.

\medskip

\noindent{\bf Step 1 ($\payoffset$ is convex).}
By \autoref{lemmaImplementation}, $(\BS,\SSurplus)\in\payoffset$ if and only if there exists an ex post allocation rule $q$ and a transfer rule $t$ whose interim outcomes satisfy Conditions 1--3, with $\BS=\E[V(\type)-T(\type)]$ and $\SSurplus=\E[T(\type)-\type Q(\type)]$.\footnote{The converse part of \autoref{lemmaImplementation} shows that the interim outcomes of any algorithm and equilibrium take this form, where $q(\wtp,\type)$ is the equilibrium trade probability. For the direct part, the transfer rule $t=Tq/Q$ (with $t=0$ when $Q=0$, in which case Condition 1 implies $T=0$) turns $q$ into a direct mechanism with these interim outcomes.} Conditions 1--3 are preserved under convex combinations of rules, and the payoffs are linear in $q$; hence $\payoffset$ is convex.

\medskip

\noindent{\bf Step 2 (Every direction is supported on $K$).}
We show that, for every $(w_B,w_S)\neq(0,0)$, the maximum of $w_B\BS+w_S\SSurplus$ over $\payoffset$ is attained at a point of $K$; the following three cases are exhaustive.

\emph{Case 1: $w_B>0$ and $w_S<w_B$.} Set $\weight=1-w_S/w_B>0$, so that $w_B\BS+w_S\SSurplus=w_B(\TS-\weight\SSurplus)$. The virtual-surplus representation \eqref{eq:objective-virtual-surplus}, which applies to every algorithm and equilibrium and extends verbatim to every $\weight\ge0$, gives
\begin{equation*}
\TS-\weight\SSurplus
=\E\!\left[q(\wtp,\type)\left(\wtp-\virtual_\weight(\type)\right)\right]
\le\E\!\left[\max\{\wtp-\virtual_\weight(\type),0\}\right].
\end{equation*}
The $\weight$-algorithm attains the bound because its trade rule is $\ind\{\wtp\ge\virtual_\weight(\type)\}$.

\emph{Case 2: $w_S>0$ and $w_B\le w_S$.} Because $\BS\ge0$, $w_B\BS+w_S\SSurplus\le w_S\TS\le w_S\TS^*$, where $\TS^*\triangleq\E[\max\{\wtp-\type,0\}]$. The $0$-algorithm attains the bound: Its allocation is efficient and leaves the buyer zero surplus (see \autoref{subsec:single-optimal}), so $(\BS_0,\SSurplus_0)=(0,\TS^*)$.

\emph{Case 3: $w_B\le0$ and $w_S\le0$.} Because $\BS\ge0$ and $\SSurplus\ge0$, $w_B\BS+w_S\SSurplus\le0$, with equality at $(0,0)$.

\medskip

\noindent{\bf Step 3 ($\payoffset=\mathrm{conv}(K)$).}
Because $K\subseteq\payoffset$ holds, and $\payoffset$ is convex, $\mathrm{conv}(K)\subseteq\payoffset$. The map $\weight\mapsto(\BS_\weight,\SSurplus_\weight)$ is continuous on $[0,\infty)$ by dominated convergence, and it converges to $(0,0)$ as $\weight\to\infty$ because the trade event $\{\wtp\ge\virtual_\weight(\type)\}$ shrinks to a null set; hence $K$ and $\mathrm{conv}(K)$ are compact. If some $(b,s)\in\payoffset\setminus\mathrm{conv}(K)$ existed, the separating hyperplane theorem would give $(w_B,w_S)\neq(0,0)$ with $w_Bb+w_Ss>\max_{(x,y)\in K}(w_Bx+w_Sy)$.
This contradicts Step 2.
Hence $\payoffset=\mathrm{conv}(K)$.

\medskip

\noindent{\bf Step 4 (Pareto frontier).}
By Step 2, for any $\beta \in [0,1]$, there is some $(\BS_\weight,\SSurplus_\weight)$ with $\weight\in[0,1]$ that maximizes $\beta \BS+(1-\beta)\SSurplus$.
Moreover, the allocation rule that maximizes this objective is unique almost everywhere; payoff equivalence and the normalization $U(1)=0$ then uniquely determine the associated payoffs.
Thus, $\{(\BS_\weight,\SSurplus_\weight)\mid\weight\in[0,1]\}$ is the Pareto frontier of $\payoffset$.

\medskip

\noindent{\bf Step 5 (Equivalence with direct mechanisms).}
Let $\mathcal{M}$ denote the set of pairs $(\BS,\SSurplus)$ attainable by direct mechanisms that are incentive compatible, seller individually rational, and give the buyer a nonnegative ex ante surplus. First, the revelation principle implies $\payoffset\subseteq\mathcal{M}$.
We show $\mathcal{M}\subseteq \payoffset$.
For any mechanism in $\mathcal{M}$, let $U(c)\triangleq T(c)-cQ(c)$ denote the seller's interim profit. Incentive compatibility and the envelope formula imply $U(c)=U(1)+\int_c^1Q(x)\,\dint x$, while seller individual rationality implies $U(1)\ge0$. Averaging over seller types gives $\SSurplus=U(1)+\E[Q(\type)\typemeas(\type)/\typedens(\type)]$.
In Case 1, with $\weight=1-w_S/w_B>0$, it follows that $\TS-\weight\SSurplus=\E[q(\wtp,\type)(\wtp-\virtual_\weight(\type))]-\weight U(1)\le\E[\max\{\wtp-\virtual_\weight(\type),0\}]$, where the inequality uses $U(1)\ge0$ and $q(\wtp,\type)\in[0,1]$. The $\weight$-algorithm attains this bound. Cases 2 and 3 use only $\BS\ge0$, $\SSurplus\ge0$, and $\TS\le\TS^*$, so their arguments apply unchanged to mechanisms in $\mathcal{M}$. Thus, every directional bound over $\mathcal{M}$ is attained at a point of $K\subseteq\payoffset\subseteq\mathcal{M}$. The separation argument of Step 3 gives $\mathcal{M}\subseteq\mathrm{conv}(K)=\payoffset$, and hence $\mathcal{M}=\payoffset$.
\end{proof}

\subsection{Closed-Form Payoff Set in the Uniform Example}\label{app:uniform-payoff-set}

This appendix derives the payoff set for the uniform setting ($\wtp\sim U[0,1]$, $\type\sim U[0,1]$) depicted in \autoref{fig:payoff-set}. Under the $\weight$-algorithm, trade occurs if and only if $\wtp\ge(1+\weight)\type$, and direct integration gives
\begin{equation*}
\BS_\weight=\frac{\weight}{3(1+\weight)^2},
\qquad
\SSurplus_\weight=\frac{1}{6(1+\weight)^2},
\qquad
\TS_\weight=\frac{1+2\weight}{6(1+\weight)^2}.
\end{equation*}
Eliminating $\weight$ shows that the $\weight$-algorithm outcomes trace the curve $\BS=\sqrt{2\SSurplus/3}-2\SSurplus$. Because the right-hand side is concave in $\SSurplus$ and vanishes at $\SSurplus=0$ and $\SSurplus=1/6$, the convex hull in \autoref{prop:payoff-set} is the region between the curve and the vertical axis:
\begin{equation*}
\payoffset=\left\{(\BS,\SSurplus)\;\middle|\;0\le\SSurplus\le\frac{1}{6},\quad0\le\BS\le\sqrt{\frac{2\SSurplus}{3}}-2\SSurplus\right\}.
\end{equation*}

\subsection{Proof of \autoref{propStrictImprovabilityBSTS}}\label{proofpropStrictImprovabilityBSTS}
\begin{proof}
For $\weight>0$, \autoref{prop:competing-sellers} shows that the optimal algorithm allocates according to weighted virtual surplus.
Because $\virtual_{\weight,j}(1)>1$, each seller has a cutoff $\overline{\type}_{\weight,j}<1$ above which the seller is inactive, so those types do not trade under the optimal algorithm.

In the no-algorithm Bertrand benchmark, every cost type below the upper support trades with strictly positive probability.
Indeed, suppose seller $j$ posts $p = \frac{1+\type_j}{2}$.
Given that values and costs are independent and have full support on $[0,1]$, there is a positive probability event such that $\wtp_j - p > \varepsilon$ and $\wtp_k < \varepsilon$ for every $k\not= j$.
On this event, seller $j$ sells to the buyer at price $p$ and earns a strictly positive profit.
Because seller $j$ can always earn a positive profit in this way, she must face a positive transaction probability at any $\type_j< 1$ in any equilibrium.
Thus the Bertrand allocation differs from the weighted virtual-surplus allocation on a positive-measure set of types.

Under the maintained regularity and continuity assumptions, the weighted virtual-surplus allocation is the optimizer of $\Wobj_\weight$. Hence the Bertrand benchmark cannot attain the optimal value and is improvable by an algorithm.
\end{proof}

\subsection{Proof of \autoref{claim:weighted-ratio-monotonicity}}\label{secAppendixWeightedRatioMonotonicity}

\begin{proof}
Fix $\weight\in(0,1]$ and define $h(c)=c+\weight\frac{F(c)}{f(c)}$ and $W(x)=v-h(x)$.
Let $\dint\mu_J(x)=J(1-F(x))^{J-1}f(x)\,\dint x$ be the distribution of the lowest cost among $J$ sellers.
Because $F/f$ is strictly increasing and nonnegative on $(0,1)$, $h$ is strictly increasing. Also, $\lim_{c\to0^+}F(c)/f(c)=0$: if the limit were strictly positive, integrating $f/F$ from $c$ to $1$ would contradict $F(c)\to0$. Hence $h(0^+)=0$.
Let $a=h^{-1}(v)$.
Then $0<a<v$, and $W(x)>0$ on $(0,a)$ while $W(x)<0$ on $(a,1]$.

Under the $\weight$-optimal algorithm, the weighted objective is
\begin{equation}
N_J(v)=\E\!\left[\left(v-h(c_{(1:J)})\right)_+\right]
=
\int_0^a W(x)\,\dint\mu_J(x).
\label{eq:weighted-NJ-new}
\end{equation}
Under Bertrand competition, by revenue equivalence, the weighted objective is
\begin{equation*}
D_J(v)
=
\weight\,\E\!\left[\left(v-c_{(2:J)}\right)_+\right]
+
(1-\weight)\,\E\!\left[\left(v-c_{(1:J)}\right)_+\right].
\end{equation*}
An integration-by-parts calculation gives
\begin{equation}
D_J(v)
=
\int_0^v W(x)\,\dint\mu_J(x).
\label{eq:weighted-DJ-new}
\end{equation}

Define $M_J(v)=\int_a^v(-W(x))\,\dint\mu_J(x)$.
Then \eqref{eq:weighted-NJ-new} and \eqref{eq:weighted-DJ-new} imply $D_J(v)=N_J(v)-M_J(v)$.
The maintained condition $D_J(v)>0$ implies $N_J(v)>M_J(v)$, and the ratio of the two objectives is
\begin{equation*}
R(J,\weight)
=
\frac{N_J(v)}{D_J(v)}
=
\frac{1}{1-M_J(v)/N_J(v)}.
\end{equation*}
Thus, it suffices to show that $M_J(v)/N_J(v)$ decreases with $J$.

The measures $\mu_J$ satisfy $\dint\mu_{J+1}(x)=\frac{J+1}{J}(1-F(x))\,\dint\mu_J(x)$.
Let $\Gamma_J=\frac{J+1}{J}(1-F(a))$.
Because $F$ is strictly increasing, the factor $\frac{J+1}{J}(1-F(x))$ is strictly larger than $\Gamma_J$ on $(0,a)$ and strictly smaller than $\Gamma_J$ on $(a,v)$.
Using the positivity of $W$ on $(0,a)$ and of $-W$ on $(a,v)$ gives
\begin{align*}
N_{J+1}(v)
&=
\frac{J+1}{J}\int_0^a(1-F(x))W(x)\,\dint\mu_J(x)
>
\Gamma_J N_J(v),\\
M_{J+1}(v)
&=
\frac{J+1}{J}\int_a^v(1-F(x))(-W(x))\,\dint\mu_J(x)
<
\Gamma_J M_J(v).
\end{align*}
Hence
\begin{equation*}
\frac{M_{J+1}(v)}{M_J(v)}
<
\Gamma_J
<
\frac{N_{J+1}(v)}{N_J(v)}.
\end{equation*}
Cross-multiplication yields $\frac{M_{J+1}(v)}{N_{J+1}(v)}<\frac{M_J(v)}{N_J(v)}$.
Therefore $R(J,\weight)$ is strictly decreasing in $J$.
The same bounds also imply $D_{J+1}(v)=N_{J+1}(v)-M_{J+1}(v)>\Gamma_J\left(N_J(v)-M_J(v)\right)>0$, so the ratio is well defined for all subsequent $J$.

Finally, $c_{(1:J)}\to0$ and $c_{(2:J)}\to0$ almost surely as $J\to\infty$.
Because the relevant payoffs are bounded, bounded convergence gives $N_J(v)\to v$ and $D_J(v)\to v$.
Thus $\lim_{J\to\infty}R(J,\weight)=1$.
\end{proof}

\subsection{Proof of \autoref{propMergerBSTS}}\label{proofpropMergerBSTS}
\begin{proof}

Fix $\weight\in(0,1]$. By \autoref{prop:competing-sellers} and the expression \eqref{eq:objective-virtual-surplus}, the pre-merger maximal value of $\Wobj_\weight$ equals $W^*\triangleq\E[\max\{0,\wtp-\virtual_\weight(\type_{(1)})\}]$, where $\type_{(1)}\triangleq\min_{j\in\Jset}\type_j$.

As in \cite{loertscher2019merger,loertscher2022incomplete}, we model the merged entity as a single seller $M$ whose cost is $\type_M=\min\{\type_1,\type_2\}$. The post-merger market is then the model of \autoref{sec:model} with sellers $\{M,3,\ldots,J\}$, where $\type_M$ has distribution $F_M=1-(1-F)^2$ with density $f_M=2f(1-F)$, positive on $[0,1)$. For every $\type\in[0,1)$, the merged entity's weighted virtual cost \eqref{eq:alpha-virtual-general} satisfies
\begin{equation}\label{eq:merger-virtual-identity}
\virtual_{\weight,M}(\type)
\;=\;
\type+\weight\,\frac{F_M(\type)}{f_M(\type)}
\;=\;
\virtual_\weight(\type)+\weight\,\frac{F(\type)^2}{2f(\type)\bigl(1-F(\type)\bigr)}
\;\ge\;
\virtual_\weight(\type).
\end{equation}

Note that the expression \eqref{eq:objective-virtual-surplus} applies to any algorithm and equilibrium of the post-merger market, even though $\virtual_{\weight,M}$ need not satisfy \autoref{ass:regular-virtual-costs}. 
Thus, every post-merger algorithm and equilibrium attains at most
\begin{equation}\label{equationBound}
\E\!\left[\max\left\{0,\;\wtp-\virtual_{\weight,M}(\type_M),\;\max_{j\ge 3}\bigl(\wtp-\virtual_\weight(\type_j)\bigr)\right\}\right].
\end{equation}
Draw $(\type_1,\ldots,\type_J)$ as in the pre-merger market and set $\type_M=\min\{\type_1,\type_2\}$, which induces the post-merger joint distribution. Because $\virtual_\weight$ is increasing, \eqref{eq:merger-virtual-identity} implies, pointwise,
$\wtp-\virtual_{\weight,M}(\type_M)\le\wtp-\virtual_\weight(\type_M)=\max_{j\in\{1,2\}}\{\wtp-\virtual_\weight(\type_j)\}$. 
Therefore, \eqref{equationBound} is at most $W^*$, i.e., the merger weakly lowers the maximal value of $\Wobj_\weight$.
\end{proof}

\subsection{Proof of \autoref{propositionComparison}}\label{proofpropositionComparison}
\begin{proof}
Under the $\weight$-optimal algorithm, trade occurs if and only if $v\ge\virtual_\weight(c)$. Because $f/F$ is strictly decreasing, $F/f$ is strictly increasing, and for any $c_1<c_2$,
\begin{equation}\label{difference}
\virtual_\weight(c_2)-\virtual_\weight(c_1)
=(c_2-c_1)+\weight\left[
\frac{F(c_2)}{f(c_2)}-\frac{F(c_1)}{f(c_1)}
\right]
>c_2-c_1.
\end{equation}

Under the myopic algorithm, let $p^m(c)$ denote the smallest monopoly price of a seller with cost $c$, so trade occurs if and only if $v\ge p^m(c)$. For the virtual value $r(v)=v-(1-G(v))/g(v)$, this price is the generalized inverse
\begin{equation*}
p^m(c)=\inf\{v\in(0,1]:r(v)\ge c\},
\end{equation*}
Because $(1-G)/g$ is strictly decreasing, we have $r(v_2)-r(v_1)>v_2-v_1$ for any $0<v_1<v_2<1$ and profit rises below and falls above the cutoff. 

We show $0\le p^m(c_2)-p^m(c_1)\le c_2-c_1$ if $c_2>c_1$.
To verify the second inequality, write $p_i=p^m(c_i)$ and suppose $p_1<p_2$. For any $p_1<x<y<p_2$, the definition of the generalized inverse gives $c_1\le r(x)<r(y)<c_2$. Moreover, the same logic as inequality \eqref{difference} implies $y-x<r(y)-r(x)$. Hence, $y-x<c_2-c_1$. Letting $x\downarrow p_1$ and $y\uparrow p_2$ yields $p^m(c_2)-p^m(c_1)\le c_2-c_1$.
Combining this with \eqref{difference}, we obtain $\virtual_\weight(c_2)-\virtual_\weight(c_1)> p^m(c_2)-p^m(c_1)$.

If $(c,v)\in\Trade^\weight\setminus\Trade^m$, then $\virtual_\weight(c)\le v<p^m(c)$, whereas if $(c',v')\in\Trade^m\setminus\Trade^\weight$, then $p^m(c')\le v'<\virtual_\weight(c')$. Thus, strict monotonicity of the threshold difference implies $c<c'$. Because $p^m$ is nondecreasing, $v<p^m(c)\le p^m(c')\le v'$, so $v<v'$.
\end{proof}

\subsection{Proof of \autoref{propositionComparisonOligopoly}}\label{proofpropositionComparisonOligopoly}
\begin{proof}
With $\typemeas(\type)=1-(1-\type)^x$ and $\typedens(\type)=x(1-\type)^{x-1}$, the common weighted virtual cost is
\begin{equation*}
\virtual_\weight(\type)=\type+\weight\frac{\typemeas(\type)}{\typedens(\type)}=\type+\frac{\weight}{x}\left[(1-\type)^{1-x}-(1-\type)\right],
\end{equation*}
with derivative $\virtual_\weight'(\type)=1+\frac{\weight}{x}\left[1+(x-1)(1-\type)^{-x}\right]$ for $\type\in[0,1)$. Because $x\ge1$, the bracketed term is at least $1$, so $\virtual_\weight'(\type)\ge1+\weight/x>1$ on $[0,1)$; hence $\virtual_\weight$ is continuous and strictly increasing on $[0,1)$, with $\virtual_\weight(0)=0$ and $\lim_{\type\to1}\virtual_\weight(\type)=1+\weight$ if $x=1$ and $+\infty$ if $x>1$ (for $x=1$, $\virtual_\weight(\type)=(1+\weight)\type$). In either case the limit exceeds $1$.

By \autoref{prop:competing-sellers}, under the $\weight$-optimal algorithm and its equilibrium, for almost every profile of values and costs the recommended option maximizes $\wtp_j-\virtual_{\weight,j}(\type_j)$ over $j\in\Jzero$, with the convention $\wtp_0=\virtual_{\weight,0}=0$. Here all sellers share the common value $\wtp_j=\wtp$ and the same strictly increasing $\virtual_\weight$, so among sellers the maximum is attained by a lowest-cost seller, and the recommended option is a nondummy seller if and only if $\wtp-\virtual_\weight(m)\ge0$. Hence, up to a null set of profiles, $\Trade^\weight=\{(m,\wtp)\mid\wtp\ge\virtual_\weight(m)\}$.

Next, consider the Bertrand benchmark of \autoref{subsec:no-algorithm}. The buyer purchases the lowest-priced product whenever $\wtp\ge\min_j p_j$. Consider a symmetric equilibrium in a strictly increasing pricing strategy $p^B:[0,1]\to\mathbb R$. A seller with cost $\type$ posting price $p$ wins if and only if every rival's price exceeds $p$, and conditional on winning, sells with probability $\Pr[\wtp\ge p]=1-\wtpmeas(p)$; her expected profit is
\begin{equation*}
(p-\type)\left[1-\typemeas\!\left((p^B)^{-1}(p)\right)\right]^{J-1}(1-\wtpmeas(p)).
\end{equation*}
This coincides with the expected profit in the homogeneous-good Bertrand model of \cite{lagerlof2024bertrand} with $n=J$ firms, demand $D(p)=1-\wtpmeas(p)=(1-p)^y$, and independent private costs drawn i.i.d.\ from $\typemeas$ (his cost-interdependence parameter set to zero). 
Corollary 1 of \cite{lagerlof2024bertrand}, specialized to his demand and signal parameters $\varepsilon=y$ and $x$, gives the symmetric equilibrium strategy
\begin{equation*}
p^B(\type)=A+(1-A)\type,
\qquad
A=\frac{1}{1+y+(J-1)x}.
\end{equation*}
Because $y>0$, $x\ge1$, and $J\ge2$, $0<A<1$, so $p^B$ is strictly increasing and the lowest-cost seller posts the lowest price; hence $\Trade^B=\{(m,\wtp)\mid\wtp\ge A+(1-A)m\}$.

Define $h(m)\triangleq\virtual_\weight(m)-A-(1-A)m$ on $[0,1)$. Then $h(0)=-A<0$; $h'(m)=\virtual_\weight'(m)-(1-A)>1-(1-A)=A>0$, so $h$ is strictly increasing; and $\lim_{m\to1}h(m)=\weight>0$ if $x=1$ and $+\infty$ if $x>1$. Hence there is a unique $m^*\in(0,1)$ with $h(m^*)=0$, and $h<0$ on $[0,m^*)$ while $h>0$ on $(m^*,1)$. Take $(m,\wtp)\in\Trade^\weight\setminus\Trade^B$: then $\virtual_\weight(m)\le\wtp<A+(1-A)m$, so $h(m)<0$ and $m<m^*$. Take $(m',\wtp')\in\Trade^B\setminus\Trade^\weight$: then $A+(1-A)m'\le\wtp'<\virtual_\weight(m')$, so $h(m')>0$ and $m'>m^*$. Therefore $m<m^*<m'$, and because $1-A>0$,
\begin{equation*}
\wtp<A+(1-A)m<A+(1-A)m'\le\wtp',
\end{equation*}
so $\wtp<\wtp'$.
\end{proof}

\subsection{Proof of \autoref{propositionDataSeller-competing}}\label{section56}\label{app:segmented-construction}
Throughout this subsection and the next, we write $\virtual_j$ for $\virtual_{\weight,j}$, $\theta_j$ for $\theta_{\weight,j}$, and $p^*_j$ for $p^*_{\weight,j}$, where $\theta_{\weight,j}$ is defined in \eqref{eq:theta-alpha}. We write $\tilde\signal_j$ for seller $j$'s signal as a random variable and $\signal_j$ for a generic realization.

For an ex post allocation rule $q$, consider the following ex post versions of Conditions 1 and 3 in \autoref{lemmaImplementation}: for every $j\in\Jset$ and every $(\vecwtp,\vectype_{-j})$,
\medskip

\noindent\emph{Condition 1*.}
$q_j(\vecwtp,\type_j,\vectype_{-j})$ is nonincreasing in $\type_j$.

\smallskip

\noindent\emph{Condition 3*.}
For every $\type_j\in[0,1]$, $\int_{\type_j}^{1}q_j(\vecwtp,x,\vectype_{-j})\,\dint x
\le
(\wtp_j-\type_j)q_j(\vecwtp,\type_j,\vectype_{-j})$.

\smallskip

Condition 3* implies $q_j(\vecwtp,\type_j,\vectype_{-j})=0$ whenever $\type_j>\wtp_j$. Together with Condition 2 in \autoref{lemmaImplementation}, integrating Conditions 1* and 3* over $(\vecwtp,\vectype_{-j})$ yields Conditions 1--3 of that lemma. For a fixed information structure and equilibrium, let $\hat Q_j(\type_j)$ denote seller $j$'s trade probability after averaging over all randomness, including signals, except seller $j$'s type.

\begin{lemma}[Implementation Lemma under Market Segmentation]\label{lem:implementation-market-segmentation}
Take any direct mechanism whose allocation rule satisfies Conditions 1* and 3* and whose interim outcomes satisfy Condition 2 in \autoref{lemmaImplementation}. Fix any information structure $\info$. Then an algorithm and an equilibrium exist under $\info$ that attain the same buyer surplus, the same total surplus, and each seller $j$'s interim profit for every $\type_j$, and that induce $\hat Q_j(\type_j)=Q_j(\type_j)$ for $\typemeas_j$-almost every $\type_j$. Moreover, the buyer is willing to follow recommendations even conditional on the recommended seller's signal realization.
\end{lemma}

\begin{proof}
All signal-conditional statements below are understood to hold for almost every signal realization. On the remaining probability-zero set, let the algorithm recommend no purchase at every price; this convention does not affect any outcome.

For each $j$, $\signal_j$, and $\type_j$, define
\[
X_j(\type_j,\signal_j)
=\E\!\left[q_j(\vecwtp,\type_j,\vectype_{-j})\mid\tilde\signal_j=\signal_j\right],
\qquad
\mathcal V_j(\type_j,\signal_j)
=\E\!\left[\wtp_jq_j(\vecwtp,\type_j,\vectype_{-j})\mid\tilde\signal_j=\signal_j\right],
\]
where the expectations are over $(\vecwtp,\vecsignal_{-j},\vectype_{-j})$ conditional on $\tilde\signal_j=\signal_j$, and set
\[
\mathcal T_j(\type_j,\signal_j)
\triangleq
\type_jX_j(\type_j,\signal_j)
+\int_{\type_j}^1X_j(x,\signal_j)\,\dint x.
\]
For almost every $\signal_j$, the triple $(X_j,\mathcal T_j,\mathcal V_j)(\cdot,\signal_j)$ satisfies Conditions 1--3 in \autoref{lemmaImplementation}. Condition 1* implies that $X_j(\cdot,\signal_j)$ is nonincreasing, and the definition of $\mathcal T_j$ is Condition 2. Condition 3* gives
\begin{equation}\label{eq:cond3-signal}
\begin{aligned}
\mathcal T_j(\type_j,\signal_j)
&=
\type_jX_j(\type_j,\signal_j)
+\E\!\left[
\int_{\type_j}^1q_j(\vecwtp,x,\vectype_{-j})\,\dint x
\mathrel{\Big|}\tilde\signal_j=\signal_j
\right]\\
&\le
\E\!\left[
\type_jq_j(\vecwtp,\type_j,\vectype_{-j})
+(\wtp_j-\type_j)q_j(\vecwtp,\type_j,\vectype_{-j})
\mathrel{\Big|}\tilde\signal_j=\signal_j
\right]\\
&=\mathcal V_j(\type_j,\signal_j).
\end{aligned}
\end{equation}
Because signals are independent of costs and the direct allocation rule does not depend on signals, averaging over $\tilde\signal_j$ recovers the unconditional schedules:
\[
\E_{\signal_j}[X_j(\type_j,\cdot)]=Q_j(\type_j),
\qquad
\E_{\signal_j}[\mathcal T_j(\type_j,\cdot)]
=
\type_jQ_j(\type_j)+\int_{\type_j}^1Q_j(x)\,\dint x
=
T_j(\type_j),
\]
where the last equality is Condition 2 in \autoref{lemmaImplementation}.

For each signal realization at which the preceding conditions hold, apply the construction in the proof of \autoref{lemmaImplementation}. Define
\[
p^*_j(\type_j,\signal_j)=
\begin{cases}
\mathcal T_j(\type_j,\signal_j)/X_j(\type_j,\signal_j)
&\text{if }X_j(\type_j,\signal_j)>0,\\
2&\text{if }X_j(\type_j,\signal_j)=0.
\end{cases}
\]
The monotonicity calculation in that proof shows that $p^*_j(\cdot,\signal_j)$ is nondecreasing. For each price in its range, let
\[
C_j(p,\signal_j)
=
\{\type_j:p^*_j(\type_j,\signal_j)=p\}.
\]
Each price preimage is an interval. If it has nonempty interior, $X_j(\cdot,\signal_j)$ is constant on its interior; denote this constant by $\bar X_j(p,\signal_j)$. If the preimage is a singleton $\{c\}$, set $\bar X_j(p,\signal_j)=X_j(c,\signal_j)$. Moreover,
\[
\mathcal T_j(c,\signal_j)=p\,\bar X_j(p,\signal_j)
\]
at every interior point of a nondegenerate preimage and at the unique point of a singleton preimage. By \eqref{eq:cond3-signal} and $\wtp_j\le1$, every price with positive allocation is at most $1$, so the price $2$ is inactive. The endpoints of the nondegenerate price preimages form a countable, hence $\typemeas_j$-null, set $N_j(\signal_j)$.

For every $(\vecwtp,\vecsignal,\vecp)$ such that $p_k$ belongs to the range of $p^*_k(\cdot,\signal_k)$ for every $k$, define
\[
\alg_j(\vecwtp,\vecsignal,\vecp)
=
\E_{\vectype}\!\left[
q_j(\vecwtp,\vectype)
\mathrel{\Big|}
\type_k\in C_k(p_k,\signal_k)\text{ for every }k
\right].
\]
The conditional type distribution is the product of the distributions induced by the price cells. For a singleton cell, use the degenerate distribution on its unique type. 
If any price lies outside the corresponding range, the algorithm recommends no purchase. Feasibility of $q$ implies that the recommendation probabilities sum to at most one.

\emph{Buyer obedience.}
Fix $\signal_j$ and a price $p_j$ in the range of $p^*_j(\cdot,\signal_j)$ at which seller $j$ is recommended with positive probability. Conditional on $(\signal_j,C_j(p_j,\signal_j))$, the recommendation probability is
\[
\E_{\type_j\in C_j(p_j,\signal_j)}
[X_j(\type_j,\signal_j)]
=
\bar X_j(p_j,\signal_j)>0.
\]
Endpoints of a nondegenerate cell have zero posterior probability, while the posterior on a singleton cell is degenerate. Therefore, \eqref{eq:cond3-signal} gives
\[
\E_{\type_j\in C_j(p_j,\signal_j)}
[\mathcal V_j(\type_j,\signal_j)]
\ge
\E_{\type_j\in C_j(p_j,\signal_j)}
[\mathcal T_j(\type_j,\signal_j)]
=
p_j\bar X_j(p_j,\signal_j).
\]
Dividing by the recommendation probability shows that the expected value conditional on a recommendation is at least $p_j$, even conditional on $\signal_j$. Obedience at the buyer's coarser information follows by averaging over that signal.

\emph{Seller equilibrium.}
Let
\[
U_j(\type_j,\signal_j)
=
\int_{\type_j}^1X_j(x,\signal_j)\,\dint x.
\]
The boundary-payoff argument in the proof of \autoref{lemmaImplementation} applies signal by signal. At interior points of a nondegenerate price cell and at singleton cells,
\[
(p^*_j(\type_j,\signal_j)-\type_j)
\bar X_j(p^*_j(\type_j,\signal_j),\signal_j)
=
U_j(\type_j,\signal_j).
\]
Both sides are continuous in the type along a nondegenerate cell, so the identity extends to its boundary points. It also holds at inactive types, where both sides are zero. For any active price $p$, choose an interior representative $\hat c_j(p,\signal_j)$ when the preimage is nondegenerate and its unique type when it is a singleton. A type $\type_j$ that deviates to $p$ obtains
\[
\begin{aligned}
(p-\type_j)\bar X_j(p,\signal_j)
&=
\mathcal T_j(\hat c_j(p,\signal_j),\signal_j)
-\type_jX_j(\hat c_j(p,\signal_j),\signal_j)\\
&\le U_j(\type_j,\signal_j),
\end{aligned}
\]
where the inequality is the incentive constraint implied by the monotonicity of $X_j$ and the envelope definition of $\mathcal T_j$. The in-range inactive price $2$ and prices outside the range yield zero. Thus posting $p^*_j(\type_j,\signal_j)$ is optimal at every signal realization at which the preceding conditions hold, including for boundary types. On the remaining probability-zero set, the no-recommendation convention makes every price optimal. Together with buyer obedience, these strategies form a perfect Bayesian equilibrium.

\emph{Payoff and surplus preservation.}
By Condition 2, the signal-averaged interim profit of type $\type_j$ is
\[
\E_{\signal_j}[U_j(\type_j,\cdot)]
=
\int_{\type_j}^1Q_j(x)\,\dint x
=
T_j(\type_j)-\type_jQ_j(\type_j).
\]
Thus every seller type's interim profit is preserved. The realized signal-conditional trade probability of type $\type_j$ is
\[
\bar X_j(p^*_j(\type_j,\signal_j),\signal_j),
\]
which equals $X_j(\type_j,\signal_j)$ outside $N_j(\signal_j)$ for almost every $\signal_j$. The set of $(\type_j,\signal_j)$ for which this equality fails is jointly measurable,\footnote{We use the standard measurability convention: signal spaces are standard Borel, the information structure is a Borel stochastic kernel, all decision rules are Borel measurable, and regular conditional distributions are chosen as measurable kernels. 
Thus, $X_j$, $\mathcal V_j$, $\mathcal T_j$, and $p_j^*$ are jointly measurable. 
Then, monotonicity of $p_j^*(\cdot,\signal_j)$ implies that the boundary points of its nondegenerate price preimages form a measurable graph.} and $N_j(\signal_j)$ is $\typemeas_j$-null at each such signal. Averaging over signals gives
\[
\hat Q_j(\type_j)=Q_j(\type_j)
\quad\text{for }\typemeas_j\text{-almost every }\type_j.
\]
Consequently, expected production cost is preserved.

For almost every $(\vecwtp,\vecsignal)$, iterated expectations over the type partition induced by the price preimages give
\[
\E_{\vectype}\!\left[
\alg_j(\vecwtp,\vecsignal,p^*(\vectype,\vecsignal))
\right]
=
\E_{\vectype}[q_j(\vecwtp,\vectype)].
\]
Multiplying by $\wtp_j$, which is measurable with respect to $(\vecwtp,\vecsignal)$, and integrating shows that expected gross value is preserved. Hence total surplus is preserved. Aggregate seller profit is preserved by the type-by-type profit equality, and buyer surplus, equal to total surplus net of seller profit, is preserved as well.
\end{proof}

We now prove \autoref{propositionDataSeller-competing}.
\begin{proof}[Proof of \autoref{propositionDataSeller-competing}]
\emph{Step 1: the optimal direct mechanism and the lemma's hypotheses.}
Write
\[
\varphi_j=\wtp_j-\virtual_j(\type_j)
\quad\text{for }j\in\Jzero,
\]
with $\varphi_0=0$. The envelope and virtual-surplus argument in the proof of \autoref{prop:competing-sellers} continues to apply after averaging over signals: signal--cost independence allows a cost type to imitate another cost type's entire signal-contingent strategy. Thus every information structure, algorithm, and equilibrium satisfies
\[
\Wobj_\weight
\le
\Wobj^*_\weight
\triangleq
\E\!\left[\max_{j\in\Jzero}\varphi_j\right].
\]

Select a canonical $\weight$-optimal direct mechanism as follows. It allocates to a maximizer of $\varphi_j$ when the maximum is positive. On a zero-virtual-surplus tie with the dummy seller, it favors a real seller, and it uses a fixed priority order to break ties among real sellers. Transfers are determined by the envelope formula with the normalization that the highest-cost type earns zero profit. In particular, its interim outcomes satisfy Condition 2 in \autoref{lemmaImplementation}.

We verify Conditions 1* and 3*. Fix $j$ and $(\vecwtp,\vectype_{-j})$, and write
\[
\rho(x)=q_j(\vecwtp,x,\vectype_{-j}).
\]
Strict monotonicity of $\virtual_j$ gives a cutoff $\bar x\in[0,1]$ such that $\rho(x)=1$ for $x<\bar x$, $\rho(\bar x)\in[0,1]$, and $\rho(x)=0$ for $x>\bar x$. Hence $\rho$ is nonincreasing. If $\rho(x)>0$, then $\wtp_j\ge\virtual_j(x)\ge x$, so $\bar x\le\wtp_j$. For $\type_j<\bar x$,
\[
\int_{\type_j}^1\rho(x)\,\dint x
=
\bar x-\type_j
\le
(\wtp_j-\type_j)\rho(\type_j).
\]
For $\type_j>\bar x$, both sides are zero, and at $\type_j=\bar x$ the integral is zero while the right-hand side is nonnegative. Thus Condition 3* also holds.

\emph{Step 2: attainment and allocation.}
By \autoref{lem:implementation-market-segmentation}, under every information structure an algorithm and equilibrium exist that attain the canonical mechanism's buyer surplus, total surplus, and interim profit for every seller type. They therefore attain $\Wobj^*_\weight$ and are $\weight$-optimal. Because the direct mechanism is independent of the information structure, this proves the payoff and surplus claims in \autoref{propositionDataSeller-competing}.

It remains to identify the ex post allocation. Let $\hat q_j(\vecwtp,\vectype)$ be the equilibrium probability of trade with seller $j$, averaging over signals. The virtual-surplus representation in the proof of \autoref{prop:competing-sellers} gives
\[
\Wobj^*_\weight
=
\E\!\left[
\sum_{j\in\Jset}
\hat q_j(\vecwtp,\vectype)\varphi_j
\right]
\le
\E\!\left[\max_{j\in\Jzero}\varphi_j\right]
=
\Wobj^*_\weight.
\]
Under \autoref{ass:regular-virtual-costs}, the independent atomless costs and strict monotonicity of the virtual costs imply that the generalized-virtual-surplus maximizer is unique almost surely. Equality in the pointwise bound therefore requires the constructed equilibrium to trade with that maximizer whenever its virtual surplus is positive and not to trade when the dummy seller is the unique maximizer. Hence its ex post allocation agrees with the canonical mechanism for almost every $(\vecwtp,\vectype)$ and is the same, as a random allocation, across all information structures. Behavior on null tie sets is immaterial to this conclusion.
\end{proof}

\medskip
\noindent\emph{Consistency with the segmented price formula.}
 For every $\type_j$ and every such $\signal_j$,
\[
X_j(\type_j,\signal_j)
=
\Pr[Y_j\ge\type_j\mid\tilde\signal_j=\signal_j]
=
\Pr[\theta_j\ge\virtual_j(\type_j)\mid\tilde\signal_j=\signal_j].
\]
The tail-integral identity gives
\[
\mathcal T_j(\type_j,\signal_j)
=
\E\!\left[
Y_j\one\{Y_j\ge\type_j\}
\mid\tilde\signal_j=\signal_j
\right].
\]
Therefore, whenever $X_j(\type_j,\signal_j)>0$,
\begin{equation}\label{price3}
\begin{aligned}
p^*_j(\type_j,\signal_j)
&=
\type_j+
\frac{\int_{\type_j}^1X_j(x,\signal_j)\,\dint x}
{X_j(\type_j,\signal_j)}\\
&=
\E\!\left[
\virtual_j^{-1}(\theta_j)
\mathrel{\Big|}
\theta_j\ge\virtual_j(\type_j),\
\tilde\signal_j=\signal_j
\right].
\end{aligned}
\end{equation}
Restoring the suppressed $\weight$ subscripts, \eqref{price3} is \eqref{eq:segmented-general-price}.

\subsection{Proof of \autoref{propositionPartitionalcompeting}}\label{app:proof-price-dispersion}
Fix seller $j$ and an active type $\type_j$, and consider a fictitious situation in which seller $j$'s prior belief over $(\vecwtp, \vectype_{-j})$ is given by the conditional distribution of $(\vecwtp, \vectype_{-j})$ given $\theta_j \ge \virtual_j(\type_j)$ (calculated starting from the true prior).
We call seller $j$'s prior in this situation the fictitious prior and any posterior belief updated from the fictitious prior a fictitious posterior.
Under any information structure, the price posted by seller $j$ with type $\type_j$, which is \eqref{price3}, is linear in seller $j$'s fictitious posterior belief over $(\vecwtp, \vectype_{-j})$ (which pins down $\theta_j$) conditional on signal realization $\signal_j$.

For each $\alpha \in \{L,H\}$, let $\hat \pi^{\type_j}_\alpha$ denote the distribution of seller $j$'s signal realizations conditional on $\theta_j \ge \virtual_j(\type_j)$ under signal $\info_\alpha$, and let $\mathcal{\wtpmeas}^{\type_j}_{\alpha}$ denote the distribution of the fictitious posterior when the signal realization follows $\hat \pi^{\type_j}_\alpha$.
By definition, the transaction-price distribution of type $\type_j$ under $\info_\alpha$ is the distribution of the price \eqref{price3} with $\signal_j\sim\hat\pi^{\type_j}_\alpha$.
If $\info_H$ is Blackwell more informative than $\info_L$ for seller $j$, then seller $j$ also gains more information about $(\vecwtp, \vectype_{-j})$, so $\mathcal{\wtpmeas}^{\type_j}_{H}$ is a mean-preserving spread of $\mathcal{\wtpmeas}^{\type_j}_{L}$.\footnote{Distribution $\mathcal{\wtpmeas}_H$ of posteriors being a mean-preserving spread of $\mathcal{\wtpmeas}_L$ means that there exist belief-valued random variables $Z_H$ and $Z_L$ such that  $Z_H \sim \mathcal{\wtpmeas}_H,Z_L \sim \mathcal{\wtpmeas}_L$ and 
$\mathbb{E}(Z_H \mid Z_L) = Z_L$.}
Because the price is linear in the fictitious posterior, the mean-preserving spread relation between $\mathcal{\wtpmeas}^{\type_j}_{H}$ and $\mathcal{\wtpmeas}^{\type_j}_{L}$ implies that the price of type $\type_j$ under $\info_H$ with $\signal_j \sim \hat \pi^{\type_j}_H$ is a mean-preserving spread of that under $\info_L$ with $\signal_j \sim \hat \pi^{\type_j}_L$.
Therefore, the transaction-price distribution of each active type of seller $j$ under $\info_H$ is a mean-preserving spread of that under $\info_L$, which proves the first claim.

It remains to prove the second claim, so suppose that $\info_H$ is Blackwell more informative than $\info_L$ for every seller. The proof of \autoref{propositionDataSeller-competing}, the identity $j^*$ of the trading seller and its cost type $\type_{j^*}$ are functions of $(\vecwtp,\vectype)$ alone almost surely, so the joint distribution of $(j^*,\type_{j^*})$ conditional on trade is the same under $\info_H$ and $\info_L$. Moreover, conditional on the buyer trading with seller $j$ of type $\type_j$, the signal $\tilde\signal_j$ follows $\hat\pi^{\type_j}_\alpha$. Hence, conditional on trade, the price the buyer pays is a mixture, with weights given by the $\info$-independent distribution of $(j^*,\type_{j^*})$, of the transaction-price distributions of the trading types.
By the first claim, each component of the mixture under $\info_H$ is a mean-preserving spread of the corresponding component under $\info_L$, and the mean-preserving-spread relation is preserved under mixtures with common weights.
Therefore, the distribution of the price paid conditional on trade under $\info_H$ is a mean-preserving spread of that under $\info_L$. \hfill $\square$

\subsection{Derivations for the Uniform Segmentation Example}\label{app:uniform-segmentation}

This appendix derives the equilibrium prices in the uniform segmentation example in \autoref{sectionData}. Let $\beta_\weight=1+\weight$, so that $\virtual_\weight(\type)=\beta_\weight\type$ and $\virtual_\weight^{-1}(\wtp)=\wtp/\beta_\weight$.
Under no segmentation, an active type $\type<1/\beta_\weight$ posts
\begin{equation*}
p_N^\weight(\type)
=
\E\!\left[
\frac{\tilde\wtp}{\beta_\weight}
\;\middle|\;
\tilde\wtp\ge\beta_\weight\type
\right]
=
\frac{\type}{2}+\frac{1}{2\beta_\weight}.
\end{equation*}
Under full segmentation, the same type posts
\begin{equation*}
p_F^\weight(\type,\wtp)=\frac{\wtp}{\beta_\weight}
\end{equation*}
whenever $\wtp\ge\beta_\weight\type$.
Both structures induce trade if and only if $\wtp\ge\beta_\weight\type$, and
\begin{equation*}
\E\!\left[
p_F^\weight(\type,\tilde\wtp)
\;\middle|\;
\tilde\wtp\ge\beta_\weight\type
\right]
=
p_N^\weight(\type).
\end{equation*}
For $\weight=1$ and $\type=1/4$, these prices are $3/8$ and $\wtp/2$, respectively, and trade occurs if and only if $\wtp\ge1/2$.

\section{Other Proofs} 
\subsection{Proof of \autoref{propositionDisclosure}}\label{proofpropositionDisclosure}
\begin{proof}
Because $\wtpmeas$ is atomless and $\type_0<1$,
\[
p^*
=\type_0+
\frac{\underline{\profit}}{1-\wtpmeas(\type_0)}
\leq
\type_0+\frac{\E[\max\{\wtp-\type_0,0\}]}{1-\wtpmeas(\type_0)}
=\E[\wtp\mid\wtp\geq\type_0].
\]
The inequality holds because the buyer can always decline to purchase, so the seller's profit cannot exceed the first-best total surplus.
Thus, at $p^*$, the buyer is willing to purchase following a recommendation.
Following no recommendation, whenever that signal has positive probability (which is the case if and only if $\type_0 >0$), $\E[\wtp\mid\wtp<\type_0]<\type_0\leq p^*$, so the buyer does not purchase.

The seller's profit at $p^*$ is
$
(p^*-\type_0)[1-\wtpmeas(\type_0)] =\underline{\profit}$.
At every other price, let the buyer follow the recommendation induced by adversarial persuasion.
A deviation to $p\geq\type_0$ then yields at most
\[
(p-\type_0)[1-\wtpmeas(\hat{\wtp}(p))]
\leq\underline{\profit},
\]
while a price below $\type_0$ yields nonpositive profit.
Thus, the proposed strategies form a perfect Bayesian equilibrium.

Finally, consider any algorithm and equilibrium.
At any fixed price, the algorithm induces an information policy about $\wtp$.
By the definition of adversarial persuasion, this policy induces a purchase probability of at least $1-\wtpmeas(\hat{\wtp}(p))$.
Thus, the seller can guarantee $\underline{\profit}$, so every equilibrium satisfies $\SSurplus\geq\underline{\profit}$.
Moreover, $\TS\leq\E[\max\{\wtp-\type_0,0\}]$, and hence
\[
\Wobj_\weight
=\TS-\weight\SSurplus
\leq \E[\max\{\wtp-\type_0,0\}]-\weight\underline{\profit}.
\]
The equilibrium constructed above is efficient and gives the seller profit $\underline{\profit}$, so it attains this bound and is $\weight$-optimal.
\end{proof}

\subsection{Proof of \autoref{claimDisclosure}}\label{proofclaimDisclosure}
\begin{proof}
For every $\weight\in[0,1]$ and every $\type$, we have $\type\leq\virtual_\weight(\type)\leq\virtual_1(\type)$, which also implies $\overline c_1\leq\overline c_\weight$.
For $\type\in[0,\overline c_1]$, because $\wtp-\type\geq0$ for any $\wtp \in [\virtual_\weight(\type), \virtual_1(\type)]$,
\[
\int_0^{\virtual_\weight(\type)}
(\wtp-\type)\,\dint\wtpmeas(\wtp)
\leq
\int_0^{\virtual_1(\type)}
(\wtp-\type)\,\dint\wtpmeas(\wtp).
\]
For $\type\in(\overline c_1,\overline c_\weight]$, we instead have
\[
\begin{aligned}
\int_0^{\virtual_\weight(\type)}
(\wtp-\type)\,\dint\wtpmeas(\wtp)
\leq \int_0^1(\wtp-\type)\,\dint\wtpmeas(\wtp)=\E[\V]-\type\leq \E[\V]-\overline c_1\leq 0.
\end{aligned}
\]
Here, the first inequality uses $\type\leq\virtual_\weight(\type)\leq1$, and the last is the condition for $\weight=1$ evaluated at $\type=\overline c_1$, where $\virtual_1(\overline c_1)=1$.
Hence, the condition for $\weight=1$ implies the condition for every $\weight\in[0,1]$ over its corresponding interval $[0,\overline c_\weight]$.

Fix $\weight$ and suppose that the condition holds.
The proof of \autoref{prop:single-weight-optimal} already verifies buyer obedience following every recommendation that occurs with positive probability, including at off-path prices, and verifies seller incentives when the buyer does not purchase following a nonrecommendation.
It remains only to show that the buyer does not purchase following a nonrecommendation.

Because $\wtpmeas$ is atomless with full support and $\virtual_\weight^{-1}$ is continuous and strictly increasing, $\vst_\weight$ is continuous and strictly increasing, with $\vst_\weight(1)=\overline c_\weight$.
Therefore, $\type\mapsto\popt_\weight(\type)=\vst_\weight(\virtual_\weight(\type))$ maps $[0,\overline c_\weight]$ onto $[\vst_\weight(0),\overline c_\weight]$.

If $p\leq\vst_\weight(0)$, the algorithm recommends the product for every value, so the nonrecommendation event is empty.
If $p\in(\vst_\weight(0),\overline c_\weight)$, there is a type $\type<\overline c_\weight$ such that $p=\popt_\weight(\type)$.
Strict monotonicity of $\vst_\weight$ implies that no recommendation is equivalent to $\wtp<\virtual_\weight(\type)$.
The condition $\int_0^{\virtual_\weight(\type)}
(\wtp-\type)\,\dint\wtpmeas(\wtp)
\leq 0$ gives $\E[\V\mid\V<\virtual_\weight(\type)]\leq\type$ 
while $p \geq\type$.
Thus, we have $p \ge \E[\V\mid\V<\virtual_\weight(\type)]$, i.e., the buyer does not purchase following a nonrecommendation.

Finally, if $p\geq\overline c_\weight$, the buyer's posterior following a nonrecommendation is the prior $\wtpmeas$.
Applying the maintained condition at $\type=\overline c_\weight$ and using $\virtual_\weight(\overline c_\weight)=1$ gives $\E[\V]\leq\overline c_\weight\leq p$.
The buyer again does not purchase following a nonrecommendation.
Thus, buyer behavior, seller demand, and seller incentives at every price are the same as in the equilibrium of \autoref{prop:single-weight-optimal}.
That equilibrium remains a perfect Bayesian equilibrium and attains the gatekeeping optimum.
Because gatekeeping gives the algorithm weakly more control over trade, this optimum is an upper bound for the current environment.
Therefore, the algorithm is $\weight$-optimal.
\end{proof}

\subsection{Proof of \autoref{propAlphaOptimalSymmetric}}\label{app:seller-surplus-deterministic-values}

In this section, we provide a general characterization of the optimal algorithm for the objective \eqref{eq:seller-objective} when values are deterministic and seller-specific.
We then specialize the characterization to seller-optimal objectives and explain how \autoref{propAlphaOptimalSymmetric} follows in the common-value, i.i.d. cost benchmark.

\begin{proposition}[Deterministic Values and Seller-Surplus Objectives]\label{propAlphaOptimalHeterogeneous}
Suppose values $\wtp_j\in(0,1]$ are deterministic and can differ across sellers.
Seller costs are independent, seller $j$'s cost has distribution $F_j$ with density $f_j>0$ on $[0,1]$, and $F_j/f_j$ is strictly increasing. 
For objective \eqref{eq:seller-objective}, the $\sellerweight$-optimal algorithm implements the following allocation: seller $j$ with $\type_j>\wtp_j$ never trades; among eligible sellers, the product is allocated to the seller with the highest ironed generalized virtual type $\bar\psi_j(\type_j)$, with ties broken uniformly. If no seller is eligible, no trade occurs.
The generalized virtual type of seller $j$ is
\begin{equation*}
\psi_j(\type_j)
=
(1-\sellerweight)(\wtp_j-\type_j)
+\sellerweight\frac{F_j(\type_j)}{f_j(\type_j)},
\end{equation*}
and ironing of $\psi_j$ is applied on the cost interval $[0,\wtp_j]$.
\end{proposition}

\begin{proof}
Consider the direct-mechanism relaxation with incentive compatibility and the pointwise buyer individual-rationality constraints.
For seller $j$, buyer individual rationality and the envelope formula imply the no-trade condition
\begin{equation*}
\type_j>\wtp_j
\quad\Rightarrow\quad
Q_j(\type_j)=0.
\end{equation*}
Indeed, $\wtp_j Q_j(\type_j)\ge T_j(\type_j)$ and
\begin{equation*}
T_j(\type_j)=\type_j Q_j(\type_j)+\int_{\type_j}^1 Q_j(x)\,\dint x
\end{equation*}
imply
\begin{equation*}
(\wtp_j-\type_j)Q_j(\type_j)\ge\int_{\type_j}^1 Q_j(x)\,\dint x\ge0.
\end{equation*}
If $\type_j>\wtp_j$, this is possible only if $Q_j(\type_j)=0$.

Substituting the envelope formula into the objective and integrating by parts, the contribution of seller $j$ can be written as
\begin{equation*}
\int_0^1 \psi_j(\type)Q_j(\type)f_j(\type)\,\dint\type,
\end{equation*}
where
\begin{equation*}
\psi_j(\type)
=
(1-\sellerweight)(\wtp_j-\type)
+
\sellerweight\frac{F_j(\type)}{f_j(\type)}.
\end{equation*}
Therefore, the integrand is relevant only on $[0,\wtp_j]$.
Thus the relaxed problem is
\begin{equation*}
\max_q
\sum_{j=1}^J
\int_0^{\wtp_j}
\psi_j(\type_j)Q_j(\type_j)f_j(\type_j)\,\dint\type_j,
\end{equation*}
subject to feasibility, nonnegativity, monotonicity of each $Q_j$, and the no-trade restriction above.

For each seller $j$, let $x_j^*=F_j(\wtp_j)$.
In quantile space define
\begin{equation*}
h_j(x)=\psi_j(F_j^{-1}(x))
\quad\text{and}\quad
H_j(x)=\int_0^x h_j(r)\,\dint r
\end{equation*}
for $x\in[0,x_j^*]$.
Let $\bar H_j$ be the least concave majorant of $H_j$ on this interval, with $\bar H_j(0)=H_j(0)=0$ and $\bar H_j(x_j^*)=H_j(x_j^*)$.
The ironed generalized virtual type is
\begin{equation*}
\bar\psi_j(\type)=\bar H_j'(F_j(\type))
\end{equation*}
where the derivative exists, with the usual interval convention at nondifferentiability points.
Because $\bar H_j$ is concave, $\bar\psi_j$ is nonincreasing in $\type$.

For any feasible allocation with nonincreasing $Q_j$, write $Q_j^*(x)=Q_j(F_j^{-1}(x))$.
Integration by parts in quantile space gives
\begin{align*}
&\int_0^{x_j^*}\bar H_j'(x)Q_j^*(x)\,\dint x
-
\int_0^{x_j^*}h_j(x)Q_j^*(x)\,\dint x\\
&\qquad
=
-\int_0^{x_j^*}\left(\bar H_j(x)-H_j(x)\right)\,\dint Q_j^*(x)
\ge0,
\end{align*}
because $\bar H_j\ge H_j$ and $\dint Q_j^*\le0$.
Summing over sellers and using pointwise feasibility,
\begin{align*}
\sum_j
\int_0^{\wtp_j}\psi_j Q_j f_j\,\dint\type
&\le
\sum_j
\int_0^{\wtp_j}\bar\psi_j Q_j f_j\,\dint\type\\
&\le
\int_{[0,1]^J}
\max\left(\{0\}\cup
\left\{\bar\psi_j(\type_j):
\type_j\le\wtp_j\right\}\right)
\prod_k f_k(\type_k)\,\dint\vectype.
\end{align*}
The proposed allocation attains the second bound by maximizing the ironed virtual type pointwise among eligible sellers.
It also attains equality in the ironing inequality, because on each ironing interval of seller $j$, $\bar\psi_j$ is constant and the competitors' costs are independent of $\type_j$.

It remains to verify implementation by an algorithm.
Feasibility is immediate because at most one seller is allocated.
Incentive compatibility holds because each $\bar\psi_j$ is nonincreasing in cost, so the induced interim allocation $Q_j$ is nonincreasing.
For buyer individual rationality, if $\type_j>\wtp_j$, the no-trade condition gives $Q_j(\type_j)=0$, and monotonicity implies $T_j(\type_j)=0$.
If $\type_j\le\wtp_j$, then $Q_j(x)=0$ for $x>\wtp_j$ and
\begin{equation*}
\wtp_j Q_j(\type_j)-T_j(\type_j)
=
(\wtp_j-\type_j)Q_j(\type_j)
-
\int_{\type_j}^{\wtp_j}Q_j(x)\,\dint x.
\end{equation*}
Because $Q_j$ is nonincreasing,
\begin{equation*}
\int_{\type_j}^{\wtp_j}Q_j(x)\,\dint x
\le
(\wtp_j-\type_j)Q_j(\type_j),
\end{equation*}
so $\wtp_j Q_j(\type_j)\ge T_j(\type_j)$.
The implementation lemma then delivers the desired algorithmic implementation.
\end{proof}

\begin{corollary}[Seller-Optimal Deterministic Values]\label{corSellerOptimalHeterogeneous}
Suppose $\sellerweight=1$ and each $F_j$ is log concave, or more generally that each $F_j/f_j$ is increasing. 
The seller-optimal algorithm allocates the product to an eligible seller $j$ satisfying $\type_j\le \wtp_j$ and having the highest $\bar\psi_j$, where
\begin{equation*}
\bar\psi_j
=
\frac{1}{F_j(\wtp_j)}
\int_0^{\wtp_j}F_j(\type)\,\dint \type
=
\wtp_j-\expect[\type_j\mid \type_j\le \wtp_j].
\end{equation*}
\end{corollary}

\begin{proof}
At $\sellerweight = 1$, $\psi_j(\type_j) = \typemeas_j(\type_j)/\typedens_j(\type_j)$. 
Log concavity of $\typemeas_j$ means $(\log \typemeas_j)' = \typedens_j/\typemeas_j$ is nonincreasing, so $\psi_j$ is nondecreasing.
The same monotonicity is exactly the condition imposed when $\typemeas_j/\typedens_j$ is increasing, so the argument below applies in either case.

Fix seller $j$ and let $x_j^*=\typemeas_j(\wtp_j)$.
In quantile space, $h_j(x) = \psi_j(\typemeas_j^{-1}(x))$ is nondecreasing, so $H_j(x) = \int_0^x h_j(r)\,\dint r$ is convex on $[0, x_j^*]$. The least concave majorant of a convex function on an interval is the chord:
$\bar{H}_j(x) = x \cdot H_j(x_j^*)/x_j^*$.
Hence $\bar{H}_j'(x) = H_j(x_j^*)/x_j^*$ is constant, and $\bar\psi_j(\type_j) = H_j(x_j^*)/x_j^*$ for all $\type_j \in [0, \wtp_j]$.

Computing the constant: with $\psi_j(\type) = \typemeas_j(\type)/\typedens_j(\type)$ and the substitution $r = \typemeas_j(\type)$,
\begin{equation*}
H_j(x_j^*) = \int_0^{x_j^*} h_j(r)\,\dint r = \int_0^{\wtp_j} \typemeas_j(\type)\,\dint\type.
\end{equation*}
Dividing by $x_j^* = \typemeas_j(\wtp_j)$ gives the first expression. For the second expression, integrate by parts:
\begin{equation*}
\int_0^{\wtp_j} \typemeas_j(\type)\,\dint\type = \wtp_j\, \typemeas_j(\wtp_j) - \int_0^{\wtp_j} \type\, \typedens_j(\type)\,\dint\type.
\end{equation*}
Dividing by $\typemeas_j(\wtp_j)$ yields $\bar\psi_j = \wtp_j - \expect[\type_j \mid \type_j \le \wtp_j]$.
\end{proof}

We now explain how \autoref{propAlphaOptimalSymmetric} follows. 
Impose $\sellerweight=1$, $\wtp_j=v$, and $F_j=F$ for every seller $j$. 
Because $F/f$ is strictly increasing, $\psi_j(\type)=F(\type)/f(\type)$ is increasing, so the ironing argument above makes $\bar\psi_j$ constant on $[0,v]$ and identical across sellers.
Thus the seller-optimal allocation trades only with sellers whose costs are at most $v$ and selects uniformly among them. 
The algorithm in \autoref{propAlphaOptimalSymmetric} implements exactly this allocation: all eligible types pool at price $v$, ineligible types choose prices above $v$ and do not trade, and the algorithm recommends uniformly among sellers who post $v$. 
Incentive compatibility is immediate from the fact that an eligible seller weakly prefers posting $v$ to receiving zero from any other price, whereas an ineligible seller would earn a negative payoff if she posted $v$ and traded. 
Buyer obedience holds because the buyer obtains zero payoff from any recommended product.

\subsection{Proof of \autoref{claimMechanismGapSellerObjective}}\label{app:seller-surplus-mechanism-gap}

\begin{claim}\label{claimMechanismGapSellerObjective}
Suppose the highest cost type is weakly greater than the highest possible value for each seller and
$$
\Pr\left(\exists j\ne k \text{ such that } \wtp_j>c_j \text{ and } \wtp_k>c_k\right)>0.
$$
Consider objective \eqref{eq:seller-objective}, and compare algorithms with direct mechanisms that can charge the buyer monetary transfers subject to ex ante buyer individual rationality.
If $\sellerweight>0$, the optimal mechanism attains a strictly higher objective value than any algorithm.
If $\sellerweight=0$, both mechanisms and algorithms attain the efficient total surplus; moreover, one can choose an optimal mechanism and an optimal algorithm that attain the same buyer surplus and seller surplus.
\end{claim}

\begin{proof}
Let
$$
\TS^*=\E\left[\max_{j\in\Jzero}\{\wtp_j-c_j\}\right]
$$
denote the efficient total surplus. For any mechanism or algorithm,
$$
(1-\sellerweight)\TS+\sellerweight\SSurplus
=\TS-\sellerweight\BS
\le \TS^*,
$$
where the inequality uses feasibility, $\TS\le \TS^*$, and ex ante buyer individual rationality, $\BS\ge0$. For algorithms, ex ante buyer individual rationality follows from buyer obedience.

We first show that a direct mechanism can attain this upper bound for any $\sellerweight>0$. Consider the efficient incentive-compatible mechanism whose seller transfers are pinned down by the envelope formula and normalized so that the highest cost type of every seller obtains zero profit. It allocates the product to a seller maximizing $\wtp_j-c_j$ whenever this maximum is positive. Because the highest cost type is weakly greater than the highest possible value, this mechanism is seller individually rational and efficient. Let $B^0$ denote the buyer's ex ante surplus under this mechanism. Now charge the buyer an additional lump-sum fee $B^0$ and distribute this fee among the sellers in a way that is independent of their reports. Seller incentive constraints and seller individual rationality are unchanged, the buyer's ex ante individual rationality constraint binds, and the mechanism attains $(\TS,\BS,\SSurplus)=(\TS^*,0,\TS^*)$. Thus the objective value is $\TS^*$.

It remains to show that no algorithm attains this value when $\sellerweight>0$. If an algorithm is inefficient, then $\TS<\TS^*$ and its objective value is strictly below $\TS^*$. Suppose instead that an algorithm induces the efficient allocation. By seller payoff equivalence, seller interim profits are pinned down by the efficient allocation and by the zero payoff of the highest cost type. Hence the buyer's ex ante surplus is uniquely determined by the efficient allocation and this normalization; it equals
$$
\E\left[\text{the second-largest element of }
\{0,\wtp_1-c_1,\ldots,\wtp_J-c_J\}\right].
$$
This expectation is strictly positive under the maintained assumption that, with positive probability, at least two sellers have values above costs. Therefore every efficient algorithm has $\BS>0$, and its objective value is $\TS^*-\sellerweight\BS<\TS^*$ when $\sellerweight>0$. Thus the optimal mechanism attains a strictly higher objective value than any algorithm.

When $\sellerweight=0$, the objective is total surplus. A direct mechanism can attain $\TS^*$ by implementing the efficient allocation, and an algorithm can also attain $\TS^*$ by the total-surplus case of the main analysis. Moreover, take the efficient direct mechanism without the additional lump-sum fee charged to the buyer. By the equivalence result for the total-surplus case in the main analysis, there is an algorithm and equilibrium that implement the same buyer surplus, total surplus, and seller profits as this mechanism. Thus, at $\sellerweight=0$, one can choose an optimal mechanism and an optimal algorithm that attain the same buyer surplus and seller surplus.
\end{proof}

\newpage

\end{document}